\pdfoutput=1
\documentclass[12pt]{elsarticle}
\usepackage{url}
\usepackage{natbib}

\usepackage[hmargin=3.0cm,vmargin=3.0cm]{geometry}
\usepackage[colorlinks,linkcolor=red,urlcolor=blue,citecolor=blue,breaklinks,bookmarksnumbered]{hyperref}

\usepackage{comment} % enables the use of multi-line comments (\ifx \fi)
\usepackage{graphicx}
\usepackage{subfig}
\usepackage{array}
\usepackage{amsmath,amssymb}
\usepackage{verbatim}
\usepackage{float}
\usepackage{tabularx}
\usepackage{color}
\newcommand\numberthis{\addtocounter{equation}{1}\tag{\theequation}}
\usepackage{amsmath,amsthm}
\usepackage[]{caption}
\usepackage{paralist}

\newcommand{\jacob}{\mathbf{J}}

\newcommand{\bmx}[0]{\begin{bmatrix}}
\newcommand{\emx}[0]{\end{bmatrix}}

\newcommand{\vect}[1]{\mathbf{#1}}
\newcommand{\vects}[1]{\boldsymbol{#1}}
\newcommand{\matr}[1]{\mathbf{#1}}

\newcommand{\vone}[0]{\vect{1}}
\newcommand{\va}[0]{\vect{a}}
\newcommand{\vb}[0]{\vect{b}}
\newcommand{\vc}[0]{\vect{c}}
\newcommand{\ve}[0]{\vect{e}}
\newcommand{\vh}[0]{\vect{h}}
\newcommand{\vv}[0]{\vect{v}}
\newcommand{\vx}[0]{\vect{x}}
\newcommand{\vw}[0]{\vect{w}}
\newcommand{\vs}[0]{\vect{s}}
\newcommand{\vf}[0]{\vect{f}}
\newcommand{\vy}[0]{\vect{y}}
\newcommand{\vg}[0]{\vect{g}}
\newcommand{\vi}[0]{\vect{i}}
\newcommand{\vo}[0]{\vect{o}}

\newcommand{\vq}[0]{\vect{q}}

\newcommand{\vu}[0]{\vect{u}}

\newcommand{\vp}[0]{\vect{p}}
\newcommand{\vr}[0]{\vect{r}}
\newcommand{\vF}[0]{\vect{F}}
\newcommand{\mW}[0]{\matr{W}}

\newcommand{\mX}[0]{\matr{X}}

\newcommand{\mU}[0]{\matr{U}}
\newcommand{\mV}[0]{\matr{V}}
\newcommand{\mA}{\matr{A}}
\newcommand{\mD}{\matr{D}}

\newcommand{\mJ}{\matr{J}}

\newcommand{\mK}{\matr{K}}
\newcommand{\mP}{\matr{P}}

\newcommand{\veta}[0]{\vects{\eta}}

\newcommand{\TT}[0]{{\vects{\theta}}}
\newcommand{\grad}[0]{\nabla}

\newtheorem{theorem}{Theorem}[section]

\begin{document}\sloppy

\begin{frontmatter}
\title{DR-RNN: A deep residual recurrent neural network for model reduction
}

\author[HWU]{J.Nagoor Kani~\corref{cor1}}
\ead{nj7@hw.ac.uk}

\author[HWU]{Ahmed H. Elsheikh}
\ead{a.elsheikh@hw.ac.uk}

\address[HWU]{
% Institue of Petroleum Engineering, \\
School of Energy, Geoscience, Infrastructure and Society,\\
Heriot-Watt University, Edinburgh, UK \\
}

\cortext[cor1]{Corresponding author}

\begin{abstract}

We introduce a deep residual recurrent neural network (DR-RNN) as an efficient model reduction technique for nonlinear dynamical systems. The developed DR-RNN is inspired by the iterative steps of line search methods in finding the residual minimiser of numerically discretized differential equations. We formulate this iterative scheme as stacked recurrent neural network (RNN) embedded with the dynamical structure of the emulated differential equations.  Numerical examples demonstrate that DR-RNN can effectively emulate the full order models of nonlinear physical systems with a significantly lower number of parameters in comparison to standard RNN architectures. Further, we combined DR-RNN with Proper Orthogonal Decomposition (POD) for model reduction of time dependent partial differential equations. The presented numerical results show the stability of proposed DR-RNN as an explicit reduced order technique. We also show significant gains in accuracy by increasing the depth of proposed DR-RNN similar to other applications of deep learning.
\end{abstract}
\end{frontmatter}

% \maketitle
%\linenumbers*[1]

\section{Introduction}
Recently, detailed numerical simulations of highly nonlinear partial differential equations representing multi-physics problems became possible due to the increased power and memory of modern computers. Nevertheless, detailed simulations remain far too expensive to be used in various engineering tasks including design optimization, uncertainty quantification, and real-time decision support. For example, Bayesian calibration of subsurface reservoirs might involve millions of numerical simulations to account for the heterogeneities in the permeability fields~\citep{Elsheikh2012,Elsheikh2013-wrr}. Model Order Reduction (MOR) provides a solution to this problem by learning a computationally cheap model from a set of the detailed simulation runs. These reduced models are used to replace the high-fidelity models in optimization and statistical inference tasks. MOR could be broadly categorized into three different classes: simplified physics based models, data-fit black box models (surrogate models)~\citep{Petvipusit2014} and projection based reduced order models commonly referred to as ROM~\citep{podwilcox}.

Physics based reduced order models are derived from high-fidelity models using approaches such as simplifying physics assumptions, using coarse grids, and/or upscaling of the model parameters.
Data-fit models are generated using regression of the high-fidelity simulation data from the input to the output~\citep{podwilcox,Petvipusit2014}. In projection based ROM, the governing equations of the system are projected into a low-dimensional subspace spanned by a small number of basis functions commonly obtained by Galerkin projection. In all projection based ROM methods, it is generally assumed that the main solution characteristics could be efficiently represented using a linear combination of only a small number of basis functions. Under this assumption, it is possible to accurately capture the input-output relationship of a large-scale full-order model (FOM) using a reduced system with significantly fewer degrees of freedom~\citep{podlassi,podberkooz}.

In projection based ROM, different methods could be used to construct the projection
bases including: Proper Orthogonal Decomposition (POD), Krylov sub-space methods, and methods based on truncated balanced realization~\citep{podlassi,tpwl}. ROM based on Proper Orthogonal Decomposition
has been widely used to model nonlinear systems~\citep{podwilcox,tpwl}. Despite the success of POD based methods, there exist a number of outstanding issues that limit the applicability of POD method as an effective reduced order modeling technique. 

One issue is related to the cost of evaluating the projected nonlinear function and the corresponding Jacobian matrix in every Newton iteration. These costs create a computational bottleneck that reduces the performance of the resulting reduced order models.
Some existing approaches for constructing a reduced order approximation to alleviate such computational bottleneck are gappy POD technique, sparse sampling, Missing Point Estimation (MPE), Best Point Interpolation Method (BPIM), Empirical Interpolation Method and Discrete Empirical Interpolation Method (DEIM)~\citep{willcox2006unsteady,barrault2004empirical,deim}. All these methods rely on interpolation schemes involving the selection of discrete spatial points for producing an interpolated approximation of the nonlinear functions. Moreover, these methods are developed especially for removing the computational complexity due to the nonlinear function in the PDE system after spatial discretization.

Another issue is related to convergence and stability of the extracted ROM. Although POD based methods decrease the calculation times by orders of magnitude as a result of reducing the state variables dimension, this reduction goes hand in hand with loss of accuracy. This may result not only in inaccurate results, but also in slow convergence and in some cases model instabilities. Slow convergence means that many iterations are needed to reach the final solution and corresponds to an increase in the computational time. Divergence is even less desirable as it produces invalid simulation results.

Artificial Neural Networks (ANN) have found growing success in many machine learning applications such as computer vision, speech recognition and machine translation~\citep{hermans2013training,he2015deep,hinton2012deep,graves2013generating}. Further, ANNs offer a promising direction for the development of innovative model reduction strategies. Neural network use in the domain of MOR is generally limited to constructing surrogate models to emulate the input-output relationship of the system based on the available simulation and experimental data~\citep{koziel2013surrogate,snoek2015scalable}. Neural networks have also been combined with POD to generate reduced order models without any knowledge of the governing dynamical systems~\citep{winter2014reduced}. One reason for developing such non-intrusive reduced order modeling methods is to address the main issues of POD-Galerkin projection ROM technique such as stability and efficient nonlinearity reduction.

Recently, Recurrent Neural Network (RNN) a class of artificial neural network where connections between units form a directed cycle have been successfully applied to various sequence modeling tasks such as automatic speech recognition and system identification of time series data~\citep{hermans2013training,he2015deep,hinton2012deep,graves2013generating}. RNN has been used to emulate the evolution of dynamical systems in a number of applications~\citep{zimmermann2012forecasting,bailordavidrecurrent} and hence has large potential in building surrogate models and reduced order models for nonlinear dynamical systems. The standard approach of modeling dynamical systems using RNN relies on three steps: (a) generating training samples from a number of detailed numerical simulations, (b) defining the suitable structure of RNN to represent the system evolution, and (c) fitting the RNN parameters to the training data. This pure data-driven approach is very general and can be effectively tuned to capture any nonlinear discrete dynamical system. However, the accuracy of this approach relies on the number of training samples (obtained by running a computationally expensive model) and on the selected RNN architecture. In addition, generic architectures might require a large number of parameters to fit the training data and thus increases the computational cost of the RNN calibration process.

Many types of recurrent neural network architectures have been proposed for modeling time-dependent phenomena~\citep{zimmermann2012forecasting,bailordavidrecurrent}. Among those, a recurrent neural network called Error Correction Neural Network (ECNN)~\citep{zimmermann2012forecasting}, that utilizes the misfit between the model output and the true output termed as model error to construct the RNN architecture. ECNN architecture~\citep{zimmermann2012forecasting} augmented the standard RNN architecture by adding a correction factor based on the model error. Further, the correction factor in ECNN was activated only during the training of RNN. In other words, ECNN takes the time series of the reference output as an input to RNN for a certain length of the time period and after that time period (i.e. in future time steps), ECNN forecasts the output without the reference output as input from the fitted model.

In the current paper, we propose a physics aware RNN architecture 
to capture the underlying mathematical structure of the dynamical system under consideration. We further extend this architecture as a deep residual RNN (DR-RNN) inspired by the iterative line search methods~\citep{bertsekas1999nonlinear, tieleman2012lecture} which iteratively find the minimiser of a nonlinear objective function. The developed DR-RNN is trained to find the residual minimiser of numerically discretized ODEs or PDEs. We note that the concept of depth in the proposed DR-RNN is different from the view of hierarchically representing the abstract input to fit the desired output commonly adopted in standard deep neural network architectures~\citep{pascanu2013construct, pascanu2013difficulty}.
The proposed DR-RNN method reduces the computational complexity from $\mathcal{O}(n^3)$ to $\mathcal{O}(n^2)$ for fully coupled nonlinear systems of size $n$ and from $\mathcal{O}(n^2)$ to $\mathcal{O}(n)$ for sparse nonlinear systems obtained from discretizing time-dependent partial differential equations.

We further combined DR-RNN with projection based ROM ideas (e.g. POD and DEIM~\citep{deim})
to produce an efficient explicit nonlinear model reduction technique with superior convergence and stability properties. Combining DR-RNN with POD/DEIM, resulted in further reduction of the computational complexity form $\mathcal{O}(r^3)$ to $\mathcal{O}(r^2)$, where $r$ is the size of the reduced order model.

The rest of this paper is organized as follows: Section 2.1 describes
dimension reduction via POD-Galerkin method followed by a discussion of DEIM in section 2.2. In Section 3, we present a brief background overview of deep neural networks (feedforward and recurrent), then we introduce the proposed DR-RNN in section 4. In section 5, we evaluate the proposed DR-RNN on a number of test cases. Finally, in Section 6 the conclusions of this manuscript are presented.

\section{Background for Model Reduction}
\label{sec-review-mor}
In this section, we first define the class of dynamical systems to be considered in this study. Following that, we present a general framework for reduced order modeling based on the concept of projecting the original state space into a low-dimensional, reduced-order space. At this point, we also discuss the computational bottleneck associated with dimensionality reduction for general nonlinear systems. Then we present the DEIM algorithm to reduce the time complexity of evaluating the nonlinear terms.

\subsection{POD-Galerikin}
We consider a general system of nonlinear differential equations of the form:
\begin{equation}
\dfrac{d\vy}{dt} = \mA~\vy + \vF(\vy)
\label{ODEFOM}
\end{equation}
where ${\vy}(\va, t)\in \mathbb{R}^n$ is the state variable at time $t$ and $\va \in \mathbb{R}^d$ is a system parameter vector. The linear part of the dynamical system is given by the matrix $\mA \in \mathbb{R}^{n\times n}$ and the vector $\vF({\vy}) \in \mathbb{R}^n$ is the nonlinear term. The nonlinear function $\vF(\vy)$ is evaluated component-wise at the $n$ components of the state variable $\vy(\va, t)$. The complete space of $\vy$ is spanned by a set of $n$ orthonormal basis vectors $\mathcal{U} = \text{span}(\vu_1~\cdots~\vu_n)$. Since $\vy$ is assumed to be attracted to a certain low dimensional subspace $\tilde{\mathcal{U}} \subset \mathcal{U}$, all the solutions of Eq.~\ref{ODEFOM} could be expressed in terms of only $r$ basis vectors ($r \ll n$) that span $\tilde{\mathcal{U}} $. The solution $\vy(\va, t)$ could then be approximated as a linear combination of these basis vectors as:
\begin{equation}
\vy = \mU_r~\tilde{\vy} + \vr_{\text{\tiny{POD}}}
\label{expansion}
\end{equation}
where $\vr_{\text{\tiny{POD}}}$ is the residual representing the part of the $\vy$ that is orthogonal to the subspace $\tilde{\mathcal{U}}$. Thus, the inner product of $\vr_{\text{\tiny{POD}}}$ with any of the basis vectors that span $\tilde{\mathcal{U}}$ is zero (i.e. $\mU_r^T~\vr_{\text{\tiny{POD}}} = 0$). The basis vectors of $\tilde{\mathcal{U}}$ are collected in the matrix $\mU_r \in \mathbb{R}^{n \times r}$ and ${\tilde{\vy}(t)} \in \mathbb{R}^r$ is the time-dependent coefficient vector. POD identifies the subspace $\tilde{\mathcal{U}}$ from the singular value decomposition (SVD) of a series of temporal snapshots of the full order system (Eq.~\ref{ODEFOM}) collected in the snapshot matrix $\mX=[(\vy_1~\cdots~\vy_T)^{\va_1}~\cdots~(\vy_1~\cdots~\vy_T)^{\va_L}] \in \mathbb{R}^{n\times (T \cdot L)}$, where $L$ is the number of different simulation runs (i.e. different initial conditions, different controls and/or different model parameters). The SVD of $\mX$ is computed as:
\begin{equation}
\mX=\mU~\Sigma~\mW^*
\end{equation}
The orthonormal basis matrix $\mU_r$ for approximating $\vy(\va, t)$ is given by the first $r$ columns of the matrix $\mU$. Substituting Eq.~\ref{expansion} into Eq.~\ref{ODEFOM} while neglecting $\vr_{\text{\tiny{POD}}}$, one gets:
 \begin{equation}
 \label{eq:poda}
  \frac{d (\mU_r~\tilde{\vy})}{dt} = \mA~\mU_r~\tilde{\vy} + \vF(\mU_r~\tilde{\vy}).
 \end{equation}
By multiplying Eq.~\ref{eq:poda} with $\mU_r^\top$, one obtains POD based ROM defined by:
 \begin{equation}
 \label{eq:pod}
  \frac{d \tilde{\vy}}{dt} = \tilde{\mA}~\tilde{\vy} + \mU_r^\top~\vF(\mU_r~\tilde{\vy})
 % \label{ODEROM}
 \end{equation}
where $\tilde{\mA} = \mU_r^\top~\mA~\mU_r$. We note that the POD-Galerkin ROM (Eq.~\ref{eq:pod}) is of reduced dimension $r \ll n$ and could be used to approximate the solution of the high-dimensional full order model (Eq.~\ref{ODEFOM}).
The computation of the first term in the right hand side of Eq.~\ref{eq:pod} involve $r^2$ operations in comparison to $n^2$ multiplications in the FOM. However, the nonlinear term $\mU_r^\top~\vF(\mU_r~\tilde{\vy})$ cannot be simplified to an $\mathcal{O}(r)$ nonlinear evaluations. 
In the following subsection, we review the Discrete Empirical Interpolation Method (DEIM) which is aimed at approximating the nonlinear term $\vF(\vy)$ in Eq.~\ref{eq:pod} using $m \ll n$ evaluations and thus rendering the solution procedure of the reduced order system independent of the high-dimensional system size $n$.
\subsection{DEIM}
\label{secdeim}
As outlined in the previous section, evaluating the nonlinear term $\vF(\mU_r~\tilde{\vy})$ in the POD-Galerkin method is still an expensive computational step, as the inner products of the full high-dimensional system is needed. The DEIM algorithm tries to reduce the complexity of evaluating the nonlinear terms in the POD based ROM (Eq.~\ref{eq:pod}) by computing the nonlinear term only at $m$ carefully selected locations and interpolate everywhere else. The nonlinear term $\vF$ in Eq.~\ref{ODEFOM} is approximated by a subspace spanned by an additional set of orthonormal basis represented as $\tilde{\mathcal{V}} = [\vv_1~\cdots~\vv_{n}]$. More specifically, a low-rank representation of the nonlinearity is computed using singular value decomposition of a snapshot matrix of the nonlinear function resulting in:
\begin{equation}
 \mX_{\vF} = \mV~\Sigma_{\vF}~\mW_{\vF}^*
\end{equation}
where, $\bf \mX_{\vF}$ is the snapshot matrix of the nonlinear function evaluated using the sample solutions $\vy(a, t)$ directly from the snapshot solution matrix $\mX$ defined in the previous section.
The $m$-dimensional basis for optimally approximating $\vF(\vy)$ is given by the first $m$ columns of the matrix $\mV$, denoted by $\mV_m$. The nonlinearity vector $\vF$ is then approximated as:
\begin{equation}
 \vF \approx \mV_m~\tilde{\vf}
 \label{deimoverdetermined}
\end{equation}
where $\tilde{\vf}(\va, t)$ is similar to $\tilde{\vy}(\va, t)$ in Eq.~\ref{expansion}.
The idea behind DEIM is to make an optimal selection of $m$ rows in $\mV_m$ such that the original over-determined system Eq.~\ref{deimoverdetermined} is approximated by an invertible system with an error as small as possible. The selection procedure described in~\citep{deim} is performed to determine the boolean matrix $\mP=[\ve_{\phi_1}~\cdots~\ve_{\phi_m}] \in \mathbb{R}^{n \times m} $ while making use of the orthonormal basis vectors $\mV_{m}=[\vv_1~\cdots~\vv_{m} ]$. The columns of the boolean matrix $\mP$ are specific columns of $n$ dimensional identity matrix~\citep{deim}. Using $\mP$, the basis interpolation of Eq.~\ref{deimoverdetermined} can be made invertible and thus solvable for $\tilde{\vf}(\va, t)$
\begin{equation}
\mP^\top \vF \approx (\mP^\top \mV_{m}) \tilde{\vf} \qquad \Rightarrow \qquad \tilde{\vf} = (\mP^\top \mV_{m})^{-1} \mP^\top \vF
% \label{deimdetermined}
\end{equation}
Using this expression of $\tilde{\vf}(\va, t)$,  the approximate nonlinear term $\vF(\mU_r~\tilde{\vy})$ in Eq.~\ref{deimoverdetermined} is formulated as:
\begin{equation}
\vF \approx \mV_{m} \cdot (\mP^\top \mV_{m})^{-1} \mP^\top \cdot \vF(a, \vy) \approx \mV_{m} (\mP^\top \mV_{m})^{-1} \cdot \vF(\mP^\top \mU_r \tilde{\vy}) = \mD \cdot \vF(\mP^\top \mU_r \tilde{\vy})
\label{deimdetermined}
\end{equation}
where $\mD= \mV_{m} (\mP^\top \mV_{m})^{-1}~$ is referred to as the DEIM-matrix. Due to the selection by $\mP$, only $m$ components of the right-side $\vF$ are needed.
In addition, for nonlinear dynamical systems, implicit time integration schemes are often used. This leads to a system of nonlinear equations that must be solved at each time step for example using Newton's method. 
At each iteration, besides the nonlinear term $\vF$, the Jacobian $\mJ_{\vF}$ of the nonlinear term must also be computed with a computational cost depending on the full order dimension $n$ during the evaluation of the reduced Jacobian matrix $\tilde{\mJ}_{\vF}$ defined by,
\begin{equation}
\begin{aligned}
\tilde{\mJ}_{\vF} = \mU_r^\top~\mJ_{\vF}(\vy)~\mU_r
\end{aligned}
\end{equation}
Similar to the approximation of $\vF$ by DEIM method, the approximation for the reduced Jacobian $\tilde{\mJ}_{\vF}$ of the nonlinear term using DEIM takes the form~\citep{deim}:
\begin{equation}
\begin{aligned}
\tilde{\mJ}_{\vF} \approx \mU_r^\top \mV_{m} (\mP^\top \mV_{m})^{-1} \mJ_{\vF}(\mP^\top \mU_r \tilde{\vy}) \mP^\top \mU_r
\end{aligned}
\end{equation}
In summary, by augmenting the standard POD formulation with DEIM, we can derive the POD-DEIM reduced order model of the form:
\begin{equation}
\label{eq:poddeim}
\frac{d \tilde{\vy}}{dt} = \tilde{\mA}~\tilde{\vy} + \mD \cdot \vF(\mP^\top \mU_r \tilde{\vy})
% \label{ODEROM}
\end{equation} 
\section{Review of standard RNN architectures}
\label{secconventionalrnn}
In this section, we briefly present the basic architecture of deep neural networks. Following that, we review standard architectures of recurrent neural networks and discuss its ability to approximate any dynamical system supported by universal approximation theorem. Then, we discuss the difficulties of training RNNs due to the vanishing gradient problem. Finally, we introduce the Long Short Term Memory (LSTM) architecture as a standard method to overcome the vanishing gradient problem in RNNs.

\subsection{Deep Feedforward Neural Network}
\label{secdeepANN}
Artificial Neural Network (ANN) is a machine learning method that expresses the input-output relationship of the form:
\begin{align}
\vy = \vy^{\text{\tiny{(ANN)}}} = \mW^\top~\phi_h(\mU^\top~\bar{\vx}) + \veta
\label{datadriven11}
\end{align}
where $\bar{\vx} = [\vx;~\vone]$, $\vx$ is the input variable, $\vy$ is the target (output) variable, $\vy^{\text{\tiny{(ANN)}}}$ is the predicted output variable obtained from ANN, $\phi_h$ is the activation function (the basis function) of the input variable, $\mU$ is the transition weight matrix, $\mW$ is the output weight matrix and $\veta$ is an unknown error due to measurement or modeling errors~\citep{bishop, hornik1989multilayer}. In the current notations, the bias terms are defined within the weight matrices by augmenting the input variable $\vx$ with a unit value~\citep{hermans2013training}.
In Eq.~\ref{datadriven11}, the target variable is modeled as a linear combination of same type of basis functions (i.e. sigmoid, perceptrons, $\tanh$ basis functions) parametrized by $\mU$. Deep ANN of depth $K$ layers is a neural network architecture of the form:
\begin{align}
\vy \approx \vy^{\text{\tiny{(ANN)}}} &=
 \mW^\top~\phi_{K-1} \left(
 \mU_{K-1}^\top~\phi_{K-2} \left(
 \cdots \phi_{1} \left(
 \mU_1^\top~\bar{\vx}
 \right)
 \right)
 \right),
\end{align}
where $\phi_{k}$ and $\mU_{k}$ are the element-wise nonlinear function and the weight matrix for the $k$th layer and $\mW$ is the output weight matrix.

\subsection{Standard Recurrent Neural Network}
\label{secstandardrnn}
Recurrent Neural Network (RNN) is a neural network that has at least one feedback connection in addition to the feedforward connections~\citep{pascanu2013difficulty}.
The standard form of RNN is a discrete dynamical system of the from~\citep{pascanu2013construct}:
\begin{align}
\vh_{t+1} &= f_h(\vh_t, \bar{\va}_{t+1}) = \phi_h (\mU^\top~\vh_t + \mV^\top~\bar{\va}_{t+1} ) \numberthis \label{rnn} \\
\vy_{t+1}^{\text{\tiny{(RNN)}}} &= \mW^\top~\vh_{t+1} \numberthis \label{rnnout}
\end{align}
where $\bar{\va}_{t+1} = [\va_{t+1};~\vone]$, $\va_{t+1}$ is the input vector at time $t+1$, 
$\phi_h$ is the activation function as defined in deep ANN and $\mU, \mV$ and $\mW$ are respectively the transition, input and output weight
matrices of standard RNN.
In Eq.~\ref{rnn}, the hidden state $\vh_{t+1}$ is estimated based on the corresponding input $\va_{t+1}$ and the hidden state $\vh_{t}$ at the previous time step.
This delayed input ($\vh_t$) can be thought of as a memory for the artificial system modelled by RNN. The order of the dynamical system expressed by RNN is the number of hidden units i.e. the size of the hidden state vector $\vh(t)$~\citep{hermans2013training}.
RNN can approximate state variables of any nonlinear difference equations as a linear combination of hidden state of standard RNN as in Eq.~\ref{rnnout} supported by the universal approximation theorem:

\begin{theorem}[Universal Approximation Theorem]
Any nonlinear dynamical system can be approximated to any accuracy by a recurrent neural network, with no restrictions on the compactness of the state space, provided that the network has enough sigmoidal hidden units \citep{RNNuk,funahashi1993approximation}.
\label{theorem}
\end{theorem}

Similar to other supervised learning methods, ANN and RNN are calibrated using training data to find the optimal parameters (neuron weights) of the ANN or RNN. 
Given a set of training sequences:
\begin{align*}
D = \lbrace ( (\va_1, \vy_1)^{\ell} \cdots (\va_t, \vy_t)^{\ell} \cdots (\va_{T}, \vy_{T})^{\ell} )\rbrace_{\ell=1}^L ,%\numberthis \label{dataset} %\\
\end{align*}
the RNN parameters $\TT = \lbrace \mU,~\mV,~\mW \rbrace$
are fitted by minimizing the function:
\begin{align}
%D = \lbrace ( (\va_1, \vy_1)^{\ell} \cdots (\va_t, \vy_t)^{\ell} \cdots (\va_{T}, %\vy_{T})^{\ell} )\rbrace_{\ell=1}^L \numberthis \label{dataset} \\
\jacob_{\text{\tiny{MSE}}}(\TT) = \frac{1}{L}\sum_{\ell=1}^L \sum_{t=1}^{T} ( \vy_{t} - \vy_{t}^{\text{\tiny{(RNN)}}} )^2 ,
\numberthis \label{mseloss}
\end{align}
where $\jacob_{\text{\tiny{MSE}}}$ known as mean square error (mse) is the average distance between the observed data $\vy_{t}$ and the RNN output ${\vy}_{t}^{\text{\tiny{RNN}}}$ across a number of samples $L$ with time dependent observations $(t=1~\cdots~T~\text{and}~\ell=1~\cdots~L)$ \citep{pascanu2013construct}.
The set of parameters $\TT$ could be estimated by backpropagating the gradient of the loss function $\jacob_{\text{\tiny{MSE}}}$ with respect to $\TT$ in time. This technique is commonly called Backpropagation Through Time (BPTT)~\citep{werbos1990backpropagation,rumelhart1988learning,pascanu2013difficulty,mikolov2014learning}.

Similar to deep learning Neural Network architectures, standard RNN has training difficulties especially in the presence of long-term
dependencies due to the vanishing and exploding gradient~\citep{pascanu2013difficulty,mikolov2014learning}. The main reason for the vanishing gradient problem is the exponential dependency of the error function gradient with respect to the weight parameters $\TT$ and the repeated multiplication of error function due to the cyclic behaviour of RNN during BPTT. This repeated multiplication causes the gradient to vanish when the absolute values of weight parameters are less than one~\citep{pascanu2013difficulty,mikolov2014learning}.
\subsection{Long Term Short Term Memory network}
\label{seclstm}
LSTM architecture~\citep{hochreiterlong} was introduced to address the aforementioned vanishing gradient problem.
The architecture of LSTM is of the form:
\begin{equation}
\begin{aligned}
\vi &= \sigma(\mU_i^\top~\vh_{t} + \mV_i^\top~\bar{\va}_{t+1}) &\qquad
\vf &= \sigma(\mU_f^\top~\vh_{t} + \mV_f^\top~\bar{\va}_{t+1}) \\
\vo &= \sigma(\mU_o^\top~\vh_{t} + \mV_o^\top~\bar{\va}_{t+1}) &\qquad
 \vg &= \tanh(\mU_g^\top~\vh_{t} + \mV_g^\top~\bar{\va}_{t+1}) \\
\vc_{t+1} &= \vc_{t}~\circ~\vf + \vg~\circ~\vi &\qquad
\vh_{t+1} &= \tanh(\vc_{t+1})~\circ~\vo
\end{aligned}
\label{lstm}
\end{equation}
where $\vi, \vf, \vo$ are the input, forget and output gates respectively, with sigmoid activation functions $\sigma$. These activation functions take the same inputs namely $\va_{t+1}, \vh_{t}$ but utilize different weight matrices $\mU, \mV$ as denoted by the different subscripts. As the name implies, $\vi, \vf$ and $\vo$ act as gates to channelize the flow of information in the hidden layer. For example, the activation of gate $\vi$ in channelizing the flow of hidden state $\vg$ is done by multiplication of $\vi$ with the hidden state value $\vg$~\citep{lipton2015critical, de2015survey}.
Input gate $\vi$ and forget gate $\vf$ decides the proportion of hidden state's internal memory $\vc_t$ and the proportion of $\vg$ respectively to update $\vc_{t+1}$. Finally, the hidden state $\vh_{t+1}$ is computed by the activation of the output gate $\vo$ in channelizing flow of internal memory $\vc_{t+1}$. If the LSTM has more than one hidden unit then the operator $\circ$ in Eq.~\ref{lstm} is an element-wise multiplication operator.

\section{Physics driven Deep Residual RNN}
\label{secmethodology}
General nonlinear dynamical systems (as formulated by Eq.~\ref{ODEFOM}) are often discretized using implicit time integration schemes to allow for large time steps exceeding the numerical stability constraints~\citep{pletcher2012computational}. This leads to a system of nonlinear residual equations depending on utilized the time stepping method. For example, the residual equation obtained from implicit Euler time integration scheme at time step $t$ takes the form:
\begin{equation}
\vr_{t+1} = \vy_{t+1} - \vy_t - \Delta t~\mA~\vy_{t+1} - \Delta t~\vF(\vy_{t+1})
\label{timeresidualEuler}
\end{equation}
To be noted, the residual equation of ROM (Eq.~\ref{eq:pod} and Eq.~\ref{eq:poddeim}) takes a similar form to Eq.~\ref{timeresidualEuler} which is solved at each time step to minimze the residual using Newton's method. In addition, performing parametric uncertainty propagation requires solving a large number of realizations, in which, each forward realization of the model may involve thousands of time steps, therefore, requiring to perform a very large number of nonlinear iterations. To alleviate this computational burden, we introduce a computationally efficient deep RNN architecture which we denote as deep residual recurrent neural network (DR-RNN) to reflect the physics of the dynamical systems.

DR-RNN iteratively minimize the residual equation (Eq.~\ref{timeresidualEuler}) at each time step by stacking $K$ network layers. The architecture of DR-RNN is formulated as:
\begin{equation}
\begin{aligned}
\vy^{(k)}_{t+1} &= \vy^{(k-1)}_{t+1} - \vw~\circ~\phi_h(\mU~\vr^{(k)}_{t+1}) &\qquad\text{for}~k = 1, \\
\vy^{(k)}_{t+1} &= \vy^{(k-1)}_{t+1} - \frac{\eta_k}{\sqrt{G_k+\epsilon}}~\vr^{(k)}_{t+1} &\qquad\text{for}~k > 1,
\label{DR-RNNequation}
\end{aligned}
\end{equation}
where $\mU, \vw, \eta_k$ are the training parameters of DR-RNN, $\phi_h$ is an activation function ($ \tanh$ in the current study), the operator $\circ$ in Eq.~\ref{DR-RNNequation} denotes an element-wise multiplication operator,  $\vr^{(k)}_{t+1}$ is the residual in layer $k$ obtained by substituting $\vy_{t+1} = \vy^{(k-1)}_{t+1}$ into Eq.~\ref{timeresidualEuler} and $G_k $ is an exponentially decaying squared norm of the residual defined as:
\begin{equation}
G_{k} = \gamma~\Vert\vr^{(k)}_{t+1}\Vert^2 + \zeta~G_{k-1}
\label{rmspropDR-RNN}
\end{equation}
where $\gamma, \zeta$ are fraction factors and $\epsilon$ is a smoothing term to avoid divisions by zero.
In this formulation, we set $\vy^{(k=0)}_{t+1} = \vy_t$.
The DR-RNN output at each time step is defined as:
\begin{equation}
\vy_{t+1}^{\text{\tiny{(RNN)}}} = \mW^\top \vy_{t+1}^K % = \mI^\top \vy_{t+1}^K,
\end{equation}
where $\mW$ is a weight matrix that could be optimized during the DR-RNN training process. However, in all our numerical test cases $\mW$, was excluded from the training process and is set as a constant identity matrix. The update equation for $k > 1$ in Eq.~\ref{DR-RNNequation} is inspired by the rmsprop algorithm~\citep{tieleman2012lecture} which is a variant of the steepest descent method. In rmsprop, the parameters are updated using the following equation:
 \begin{equation}
 \begin{aligned}
 G_{k} &= (1 - \gamma)~(\grad_{\TT} \jacob (\TT^{(k)}))^2 + \gamma~G_{k-1},\\
 \TT^{(k)} &= \TT^{(k-1)} - \frac{\eta}{\sqrt{G_{k} + \epsilon}}~\grad_{\TT} \jacob(\TT^{(k-1)}),
 \end{aligned}
 \label{rmsprop}
 \end{equation}
 where $G_{k}$ is an exponentially decaying average of the squared gradients of the loss function $\jacob(\TT)$, $\gamma$ is the fraction factor (usually $0.9$), $\eta$ is the constant learning rate parameter (usually $0.001$) and $\epsilon$ is a smoothing term to avoid divisions by zero.
We note that $G_{k}$ in Eq.~\ref{rmsprop} is a vector and changes both the step length and the direction of the steepest decent update vector. However, $G_{k}$ in Eq.~\ref{DR-RNNequation} is a scaler and changes only the step size to update $\vy^{(k)}_{t+1}$ in the direction of $\vr^{(k)}_{t+1}$. Furthermore, we use $G_{k}$ as a stability factor in updating $\vy^{(k)}_{t+1}$ since the update scheme in DR-RNN is explicit in time and may be prone to instability when using large time steps.

One of the main reasons to consider DR-RNN as a low computational budget numerical emulator is the way the time sequence of the state variables is updated. The dynamics of DR-RNN are explicit in time with a fixed computational budget of order $\mathcal{O}(n)$ per time step. Furthermore, DR-RNN framework has a prospect of applying DR-RNN to solve Eq.~\ref{timeresidualEuler} on different levels of time step much larger than the time step $\Delta t$ taken in Eq.~\ref{timeresidualEuler}. In other words, DR-RNN provides an effective way to solve Eq.~\ref{timeresidualEuler} for a fixed time discretization error.

\section{Numerical Results}
In this section, we demonstrate two different applications of DR-RNN as a model reduction technique. 
The first application concerns the use of DR-RNN for reducing the computational complexity from $\mathcal{O}(n^3)$ to $\mathcal{O}(n^2)$ at each time step for nonlinear ODE systems without reducing the dimension of the state variable of the system. Moreover, DR-RNN is allowed to take large time steps violating the numerical stability condition and is constrained to have time discretization error several times less than the order of large time step taken. We denote this reduction in computational complexity as temporal model reduction. The second application is focused on spatial dimensionality reduction of dynamical systems governed by a time dependent PDEs with parametric uncertainty. In this case, we use DR-RNN to approximate a reduced order model derived using a POD-Galerkin strategy. 

In section~\ref{sectemporalmodelreduction}, DR-RNN is first demonstrated for temporal model order reduction. In addition, we provide a numerical comparison against ROM based on standard recurrent neural networks architectures.  In section~\ref{secspatialmodelreduction}, we build DR-RNN to approximate POD based reduce order model (Eq.~\ref{eq:pod}) and compare the performance of DR-RNN in approximating POD based ROM against the ROM based on the POD-Galerkin and POD-DEIM methods.

\subsection{Temporal model reduction}
\label{sectemporalmodelreduction}
In this section, we conduct temporal model reduction to evaluate the performance of DR-RNN in comparison to    the standard recurrent neural networks on three test problems.
The standard recurrent neural networks used are RNN and LSTM denoted by RNN$ _m$ and LSTM$ _m$ respectively, where the subscript $m$ denotes the order of the recurrent neural network ($ m = $number of neurons in the hidden layer).
The DR-RNN is denoted by DR-RNN$_m$, where the subscript $m$ in this case denotes the number of residual layers. We also note that the order of DR-RNN is same as the order of the given dynamical equation since we rely on using the exact expression of the system dynamics. In all test cases, we utilize a $\tanh$ activation function in the standard RNN models. 

All the numerical evaluations are performed using the \textsf{keras} framework~\citep{chollet2015keras}, a deep learning python package using~\textsf{Theano}~\citep{2016arXiv160502688short} library as a backend. Further, we train all RNN models using rmsprop algorithm~\citep{tieleman2012lecture, chollet2015keras}  as implemented in \textsf{keras} with default settings.
We set the weight matrix $\mU$ of DR-RNN in Eq.~\ref{DR-RNNequation} as a constant identity matrix and do not include it in the training process.  The vector training parameter $ \vw $ in Eq.~\ref{DR-RNNequation} is initialized randomly from a zero-mean Gaussian distribution with standard deviation fixed to  0.1. The scalar training parameters $\eta_k$ in Eq.~\ref{DR-RNNequation}  are initialized randomly from the uniform distribution $\mathtt{U [0.1, 0.4]}$. We set the hyperparameters $\zeta$ and $\gamma$ in Eq.~\ref{rmspropDR-RNN} to $0.9$ and $0.1$, respectively.

\subsubsection*{Problem 1}
\begin{figure}[!h]
\centering
\includegraphics{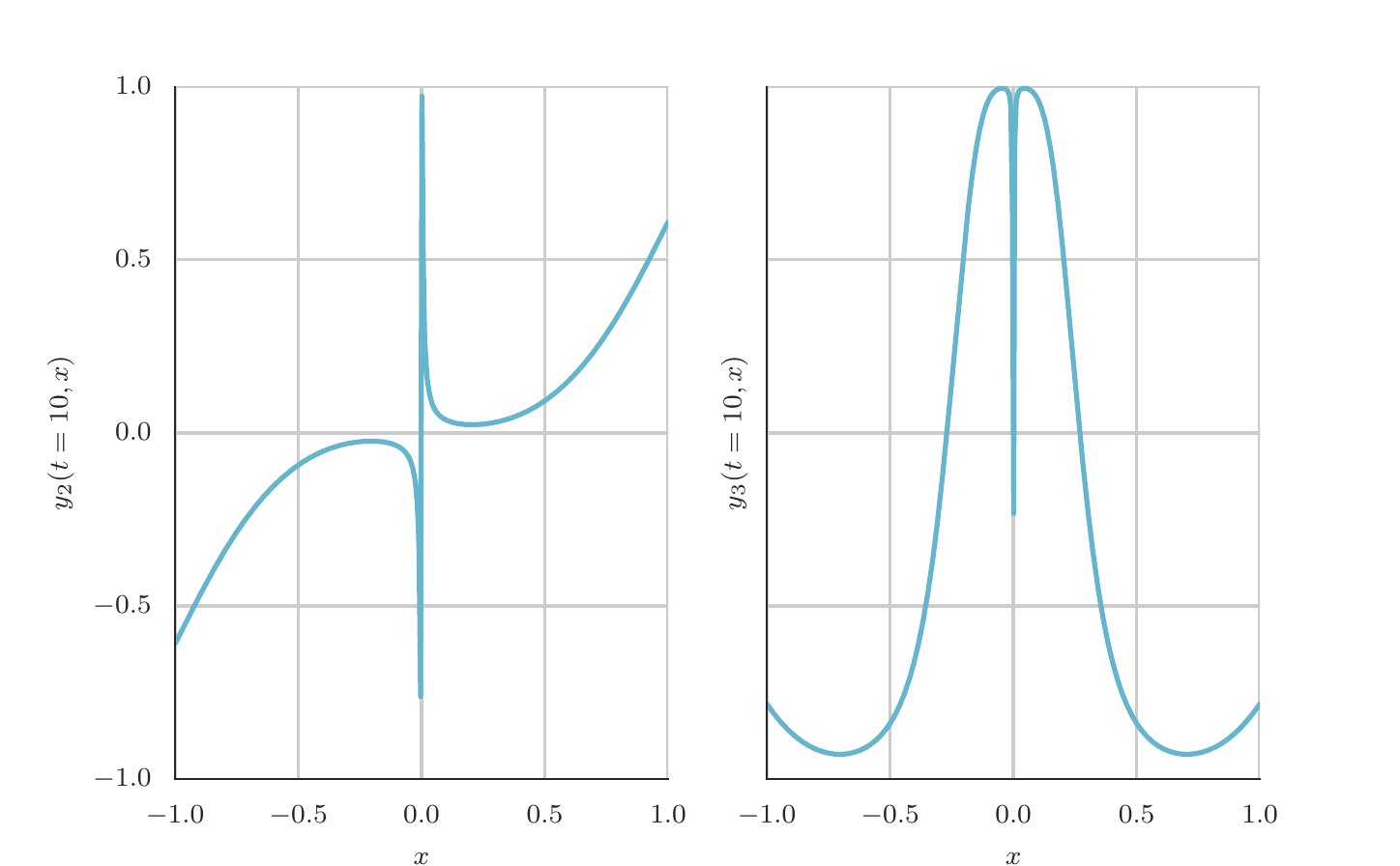}
\caption{System response versus the different values of the initial value random variable $x$. Note the jump discontinuities in the response $y_2(t=10)~\text{and}~y_3(t=10)$ at $x = 0$.}
\label{figY123}
\end{figure}

We consider a nonlinear dynamical system of order $n= 3$ defined by:
\begin{equation}
\dfrac{dy_1}{dt} = y_1~y_3, \qquad \dfrac{dy_2}{dt} = -y_2~y_3, \qquad \dfrac{dy_3}{dt} = -y_1^2 + y_2^2
\label{p1}
\end{equation}
with initial conditions $y_1(0)=1,~y_2(0)=0.1~x,~y_3(0)=0$. The input $x$ is a random variable with a uniform distribution $\mathtt{U [-1, 1]}$. Modeling this dynamical system is particularly challenging as the response has a discontinuity at the planes $y_1(0)=0$ and $y_2(0)=0$~\citep{bilionis}. Figure~\ref{figY123} shows the jump discontinuities in the response $y_2(t=10)$ and $y_3(t=10)$ versus the perturbations in the initial input $x$. A standard backward Euler method is used for 100 time steps of size $\Delta t=0.1$ and we solve the problem for 1500 random samples of $x$.
%%--------------------------------
%%--------------------------------

We train RNN parameters using data obtained from 500 random samples and the remaining runs (i.e. 1000) are used for validation.
The training is performed using a batch size of $15$ for $15$ iterations. We use 7 recurrent neural networks namely $\text{RNN}_{n}$, $\text{RNN}_{10n}$, $\text{LSTM}_{n}$, $\text{LSTM}_{10n}$, $\text{DR-RNN}_1$, $\text{DR-RNN}_2$ and $\text{DR-RNN}_4$. The performances of all 7 RNNs is evaluated based on accuracy and model complexity. Accuracy is measured using the mean square error (Eq.~\ref{mseloss}) for the training and the test data sets. Also, we show comparative plots of the probability density function (PDF) of the state variables at specific time steps.  
Model complexity is determined based on the number of parameters $d$ fitted in each RNN model.

\begin{figure}[!h]
\centering
\includegraphics[width=\linewidth]{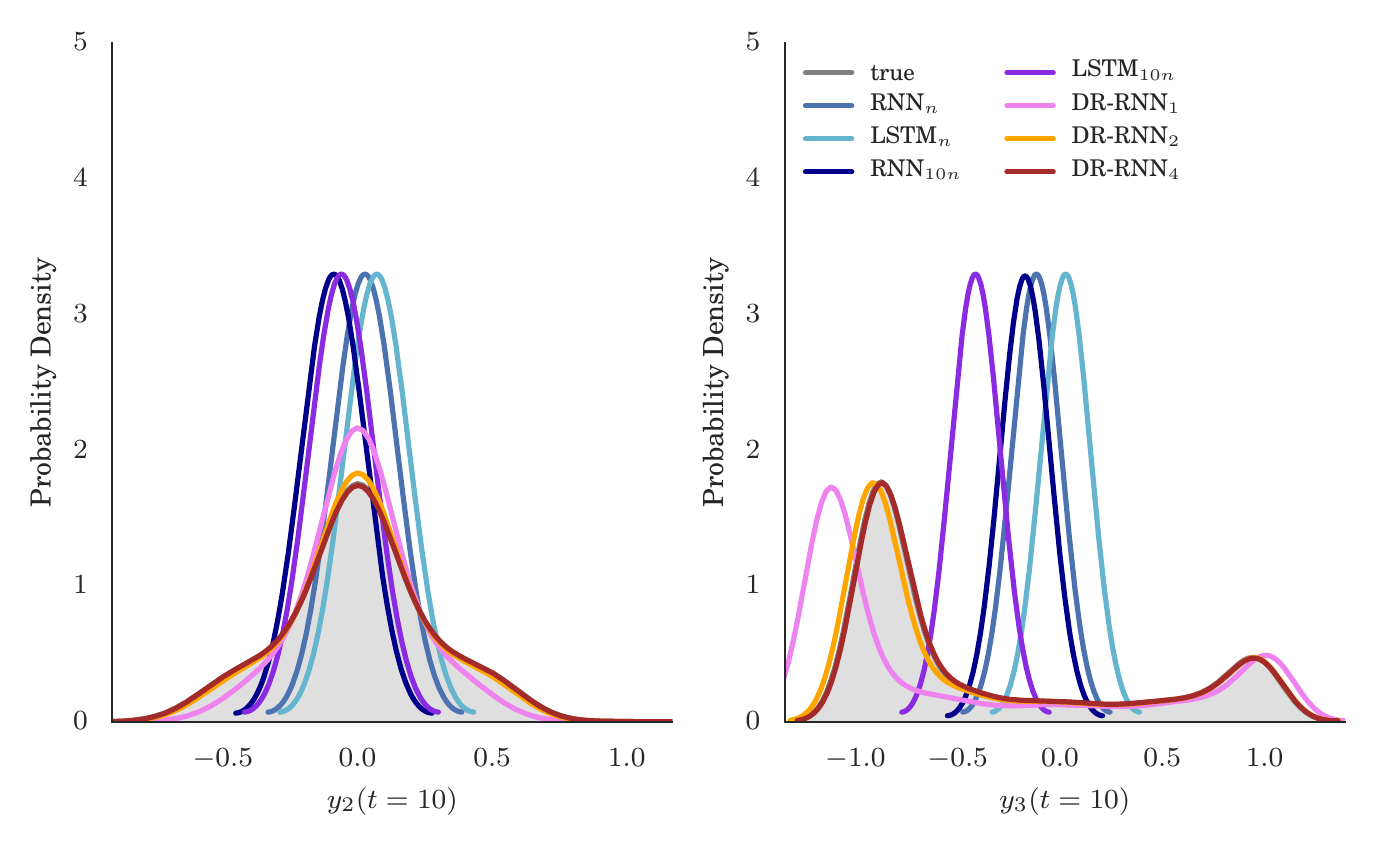}
\caption{Comparison of kernel density estimated probability density function (PDF) of $y_2(t=10)$ (left) and $y_3(t=10)$ (right) obtained from all RNN w.r.t. true PDF in problem 1. Label RNN denotes standard RNN (equation \ref{rnn}). Subscripts ($ n$ or $10n$) in the label RNN and LSTM denotes the dimension of hidden layer where $n$ is the dimension of state variable $\vy$. Subscript in the label DR-RNN denotes the number of layers $K$ in DR-RNN. The dimension of all the layers in all DR-RNN is $n$.}
\label{figY1}
\end{figure}

Figure~\ref{figY1} compares the PDF of $y_2(t=10)$ and $y_3(t=10)$ computed from all 7 RNN against the reference PDF solution. The results presented in Figure~\ref{figY1} shows that the PDF obtained from DR-RNN with residual layers closely follow the trend of the reference PDF. The mse of all RNN models and the corresponding model complexity are presented in Table~\ref{tableY1}. It is worth noticing that DR-RNN models have fewer number of parameters $d$ and hence much lower model complexity than standard RNN models. Furthermore, Table~\ref{tableY1} shows that DR-RNN with residual layers is considerably better than the standard RNN in fitting the data both in the training and testing data sets. We argue that such performance is due to the iterative update of DR-RNN output towards the desired output. However, the small differences among the models with residual layers indicates that the additional residual layers in {DR-RNN}$_{4}$ are not needed in this particular problem.

%%--------------------------------
%%--------------------------------

\begin{figure}[!h]
\centering
\includegraphics[width=\linewidth]{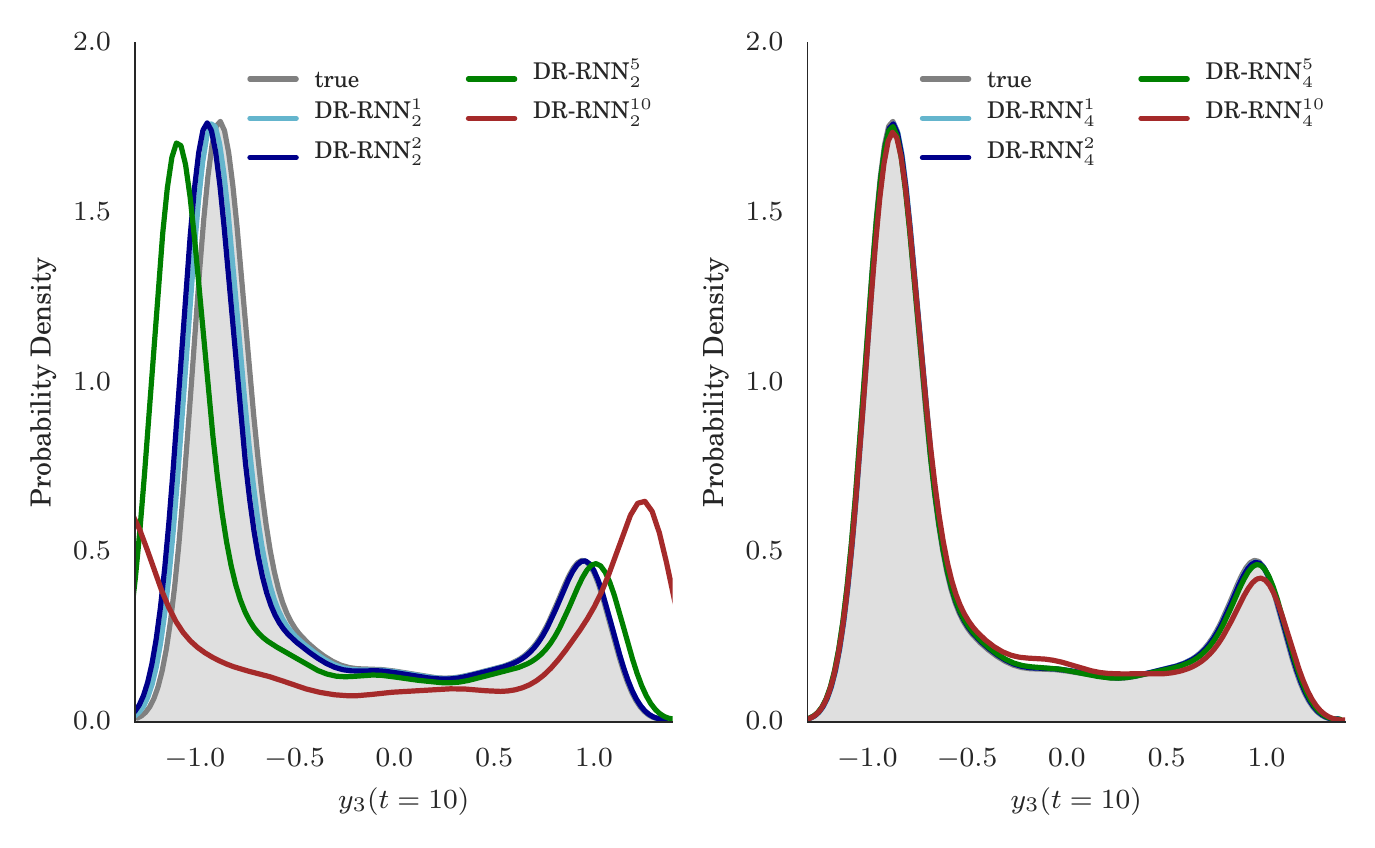}
\caption{Comparison of kernel density estimated probability density function (PDF) of $y_2(t=10)$ (left) and $y_3(t=10)$ (right) obtained from DR-RNN for different large time step size w.r.t. true PDF computed from fine step size in problem 1. Subscript in the label DR-RNN denotes the number of layers $K$ in DR-RNN. Superscript in the label DR-RNN denotes how many times the large time step bigger than the fine step size.}
\label{figY1largetimestep}
\end{figure}

%%--------------------------------
\begin{table}[!h]
\centering
\caption{Performance chart of all 7 RNN in problem 1 where $d$ is the number of parameters fitted in RNN and
mse (Eq.~\ref{mseloss}) measures the accuracy of RNN.}
\label{tableY1}
{\small
\begin{tabular}{cccccccc}
\noalign{\smallskip} \hline \hline \noalign{\smallskip
}
{\tiny Model} & {\tiny $\text{RNN}_{n}$ } & {\tiny $\text{RNN}_{10n} $} & {\tiny $\text{LSTM}_{n} $} & {\tiny $\text{LSTM}_{10n} $} & {\tiny $\text{DR-RNN}_1 $} & {\tiny $\text{DR-RNN}_2$ } & {\tiny $\text{DR-RNN}_4$ } \\
\hline
$ d$ & 33 & 84 & 1093 & 4053 & 3 & 4 & 6\\
mse train & 23$ \cdot 10^{-2}$ & 15$ \cdot 10^{-2}$ & 21$ \cdot 10^{-2}$ &15 $\cdot 10^{-2}$ & 2$ \cdot 10^{-3}$ & 4$ \cdot 10^{-5}$ & 4$ \cdot 10^{-6}$ \\
mse test & 23$ \cdot 10^{-2}$ & 15$ \cdot 10^{-2}$ & 21$ \cdot 10^{-2}$ &14 $\cdot 10^{-2}$ & 5$ \cdot 10^{-3}$ & 4$ \cdot 10^{-5}$ & 4$ \cdot 10^{-6}$ \\
\noalign{\smallskip} \hline \noalign{\smallskip
}
\end{tabular} }
\label{tableY1}
\end{table}
%%--------------------------------
We further train the DR-RNN using data sampled at time interval larger than those used in the backward Euler numerical integrator. 
For example, we train using sampled data at $\Delta t = 0.5$ resulting in 20 time samples instead of 100 time samples when using the original time step size $\Delta t = 0.1$. We analyse this experiment using DR-RNN$ _2$ and DR-RNN$ _4$ as the top performer in the last set of numerical experiments. Figure~\ref{figY1largetimestep} shows the PDF of $y_3(t=10)$ computed from DR-RNN$ _2$ and DR-RNN$ _4$ for different time step along with the PDF computed from the reference solution. As can be observed, the performance of DR-RNN$ _4$ is superior to DR-RNN$ _2$ supporting our argument on the hierarchical iterative update of the DR-RNN solution as the number of residual layer increases. In Figure~\ref{figY1largetimestep}, DR-RNN$ _2$ performed well for 2 times $\Delta t = 0.1$, while it results in large errors for 5 and 10 times $\Delta t = 0.1$ whereas DR-RNN$ _4$ performed well for all large time steps. Through this numerical experiment, we provide numerical evidence that DR-RNN is numerically stable when approximating the discrete model of the true dynamical system for a range of large time steps with small discretization errors. However, there is a limit on the time step size for the desired accuracy in the output of the DR-RNN and this limit is correlated to the number of utilized layers.
%%--------------------------------
%%--------------------------------

\subsubsection*{Problem 2}
\begin{figure}[!h]
\centering
\includegraphics[width=\linewidth]{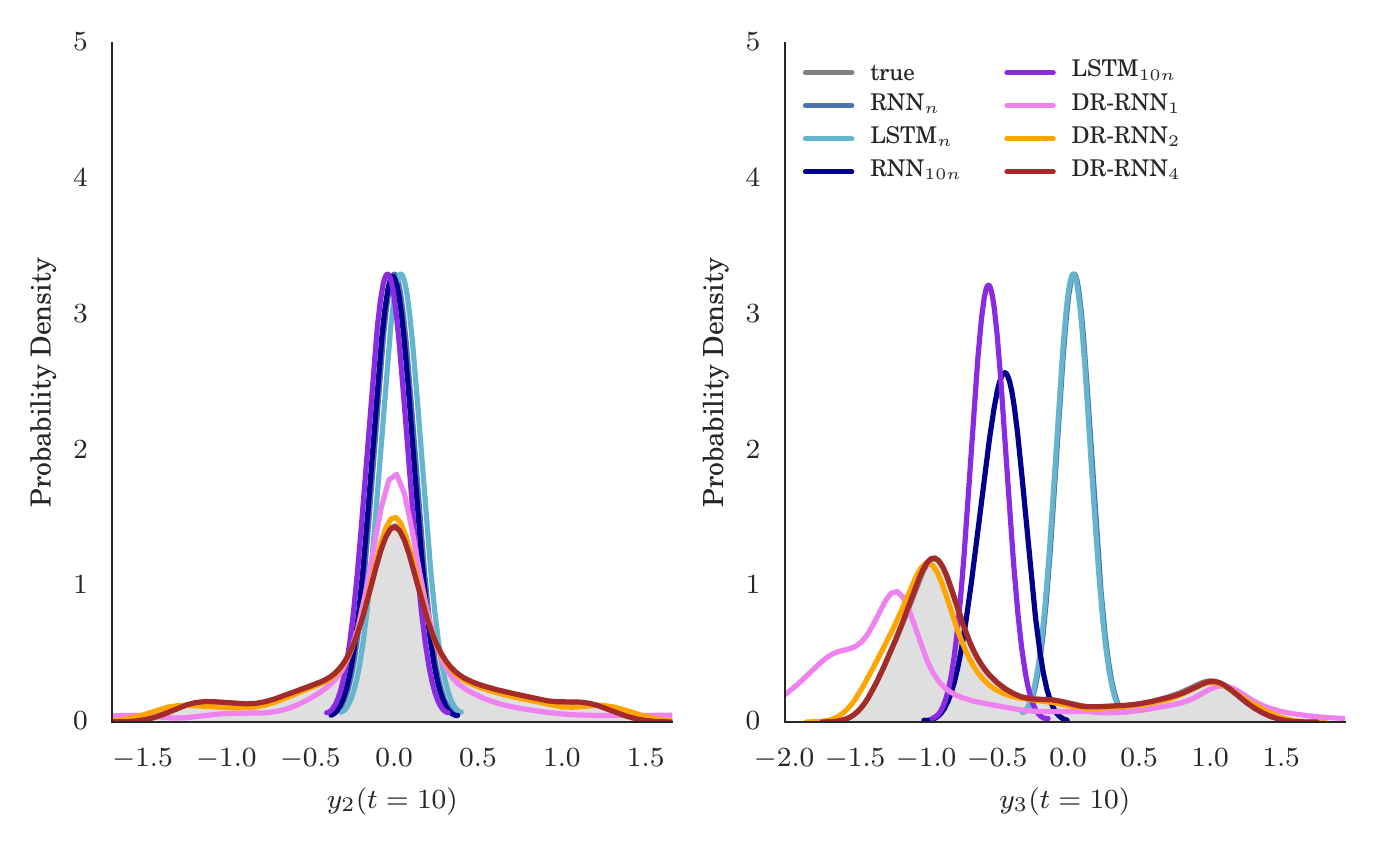}
\caption{Comparison of kernel density estimated probability density function (PDF) of $y_2(t=10)$ (left) and $y_3(t=10)$ (right) obtained from all RNN w.r.t. true PDF in problem 2. Label RNN denotes standard RNN (equation \ref{rnn}). Subscripts ($ n$ or $10n$) in the label RNN and LSTM denotes the dimension of the hidden layer where $n$ is the dimension of state variable $\vy$. Subscript in the label DR-RNN denotes the number of layers $ K $ in DR-RNN. The dimension of all the layers in all DR-RNN is $n$.}
\label{figY1Y2}
\end{figure}

The dynamical equation for problem 2 is the same as in test problem 1. However, the initial conditions are set to $y_1(0)=1,~y_2(0)=0.1~x_1,~y_3(0)=x_2$ where the stochastic dimension is increased from 1 to 2. The input random variables $x_1, x_2$ are modeled by uniform probability distribution function $\mathtt{U [-1, 1]}$.
We adopted the same procedure followed in problem 1 to evaluate the performances of the proposed DR-RNN in-comparison to the standard recurrent neural network models.
Figure~\ref{figY1Y2} shows a comparison of the PDF plot for $y_2(t=10)$ and $y_3(t=10)$ computed from all RNN models. Errors of all RNN models and the corresponding model complexity are presented in Table~\ref{tableY1Y2}. We can see the performance trend of all RNN models observed in problem 2 are similar to the trends observed in Problem 1.

%%--------------------------------
\begin{table}[!h]
\centering
\caption{Performance chart of all 7 RNN in problem 2 where $d$ is the number of parameters fitted in RNN and mse (Eq.~\ref{mseloss}) measures the accuracy of RNN.}
\label{tableY1Y2}
{\small
\begin{tabular}{cccccccc}
\noalign{\smallskip} \hline \hline \noalign{\smallskip
}
{\tiny Model} & {\tiny $\text{RNN}_{n}$ } & {\tiny $\text{RNN}_{10n} $} & {\tiny $\text{LSTM}_{n} $} & {\tiny $\text{LSTM}_{10n} $} & {\tiny $\text{DR-RNN}_1 $} & {\tiny $\text{DR-RNN}_2$ } & {\tiny $\text{DR-RNN}_4$ } \\
\hline
$ d$ & 33 & 84 & 1093 & 4053 & 3 & 4 & 6\\
mse train & 26$ \cdot 11^{-2}$ & 26$ \cdot 10^{-2}$ & 26$ \cdot 10^{-2}$ &20 $\cdot 10^{-2}$ & 2$ \cdot 10^{-2}$ & 1$ \cdot 10^{-4}$ & 2$ \cdot 10^{-6}$ \\
mse test & 26$ \cdot 11^{-2}$ & 26$ \cdot 10^{-2}$ & 26$ \cdot 10^{-2}$ &20 $\cdot 10^{-2}$ & 2$ \cdot 10^{-2}$ & 1$ \cdot 10^{-4}$ & 3$ \cdot 10^{-6}$ \\
\noalign{\smallskip} \hline \noalign{\smallskip
}
\end{tabular} }
\end{table}
%%--------------------------------

%%--------------------------------
%%--------------------------------

\begin{figure}[!h]
\centering
\includegraphics[width=\linewidth]{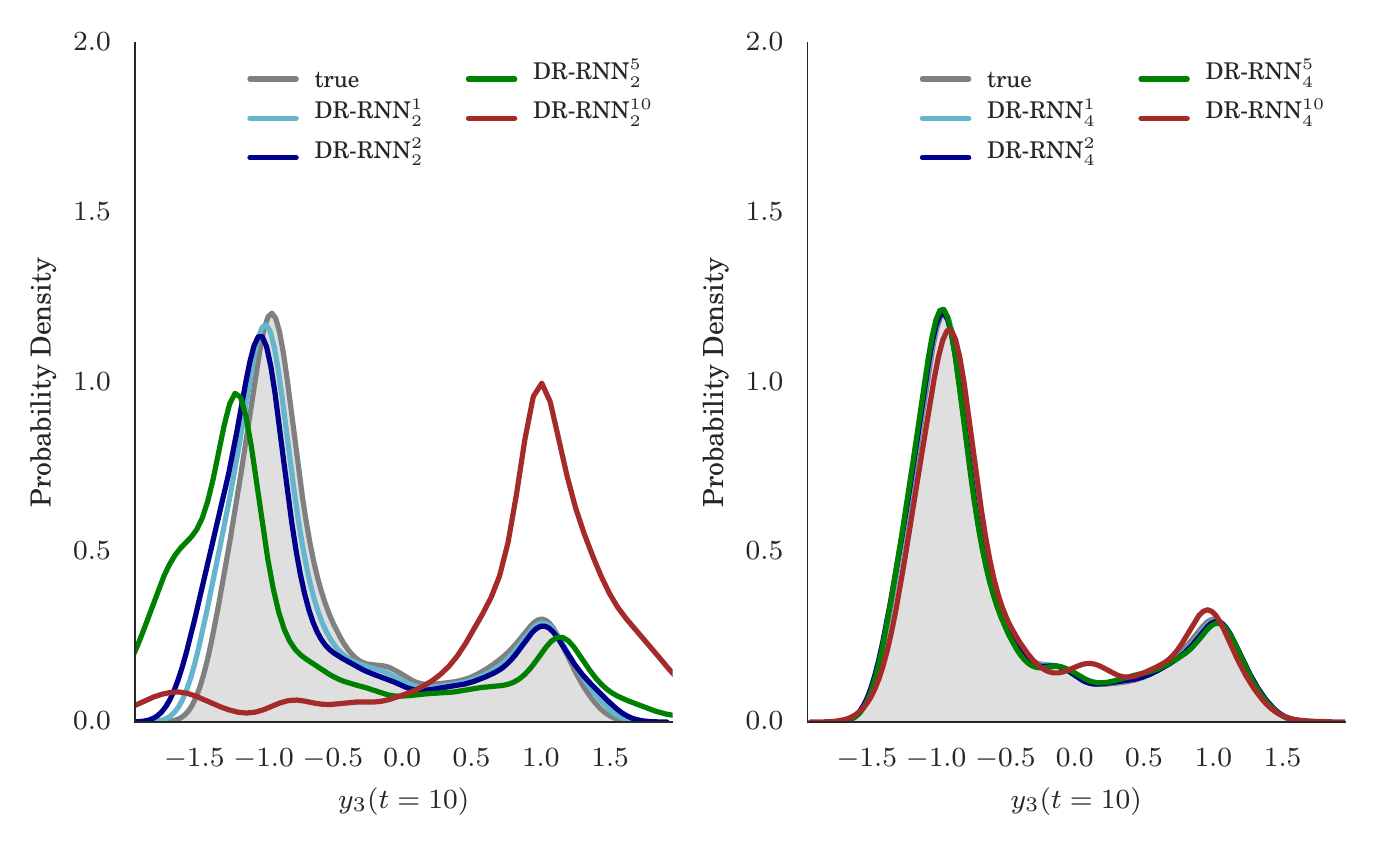}
\caption{Comparison of kernel density estimated probability density function (PDF) of $y_2(t=10)$ (left) and $y_3(t=10)$ (right) obtained from DR-RNN for different large time step size w.r.t. true PDF computed from fine step size in problem 2. Subscript in the label DR-RNN denotes the number of layers $K$ in DR-RNN. Superscript in the label DR-RNN denotes how many times the large time step bigger than the fine step size.}
\label{figY1Y2largetimestep}
\end{figure}

We follow the similar procedure adopted in problem 1 to analyze the performance of DR-RNN in taking large time step. Figure~\ref{figY1Y2largetimestep} compares the PDF of $y_3(t=10)$ computed from DR-RNN$ _2$ and DR-RNN$ _4$ for different large time steps with the PDF computed from the reference solution for the fine time step size. We observe similar performance trends of DR-RNN$_2$ and DR-RNN$_4$ to those observed in test problem 1 (Figure~\ref{figY1largetimestep}).
%%--------------------------------
%%--------------------------------

\subsubsection*{Problem 3}
\begin{figure}[!h]
\centering
\includegraphics[width=\linewidth]{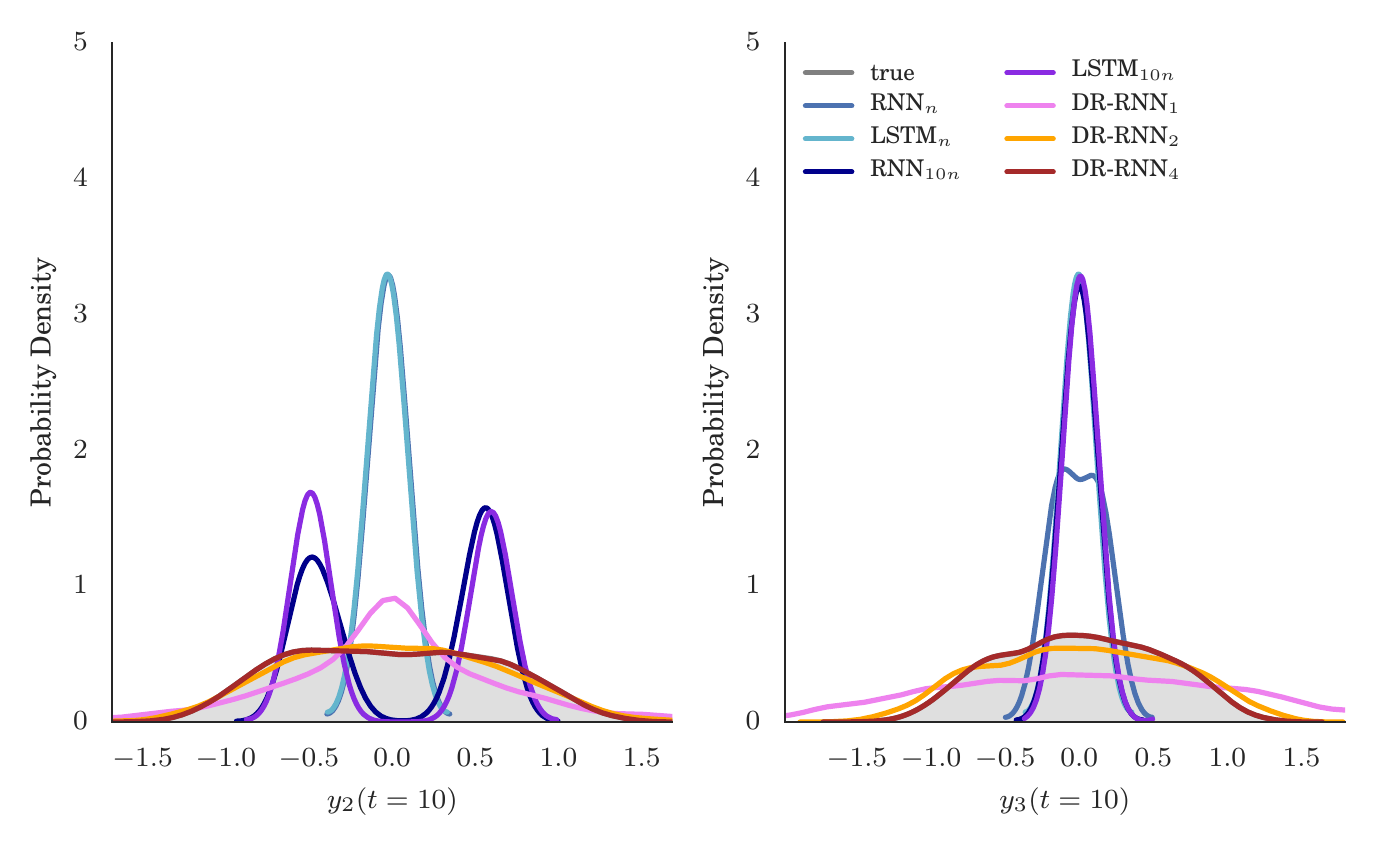}
\caption{Comparison of kernel density estimated probability density function (PDF) of $y_2(t=10)$ (left) and $y_3(t=10)$ (right) obtained from all RNN w.r.t. true PDF in problem 3. Label RNN denotes standard RNN (Eq.~\ref{rnn}). Subscripts ($ n$ or $10n$) in the label RNN and LSTM denotes the dimension of the hidden layer where $n$ is the dimension of state variable $\vy$. Subscript in the label DR-RNN denotes the number of output layers $K$ in DR-RNN. The dimension of all the layers in all DR-RNN is $n$.}
\label{figY3}
\end{figure}

The dynamical system considered in this problem is similar to problem 1 and problem 2 with further additional difficulties in the initial conditions $y_1(0)=x_1,~y_2(0)=x_2,~y_3(0)=x_3$, where $x_1, x_2, x_3 \in \mathtt{U [-1, 1]}$. Remarkably, problem 3 is rather difficult to train by RNN compared to problem 1 as the stochastic dimension in this problem is 3. We adopted the same procedure followed in problem 1 to evaluate the performances of the proposed DR-RNN in comparison to the standard recurrent neural network models. Figure~\ref{figY3} shows the PDF of $y_2(t=10)$ and $y_3(t=10)$ computed from all RNN. Errors of all RNN models and their model complexity are presented in Table~\ref{tableY3}. Performance ranking of all 7 RNN models remain similar to Problem 1 and Problem 2 in spite of the increased stochastic dimension. 
More specifically, from Table~\ref{tableY3}, we notice a decreases in mse as the number of network layers in DR-RNN increases. 

%%--------------------------------
\begin{table}[!h]
\centering
\caption{Performance chart of all 7 RNN in problem 3 where $d$ is the number of RNN parameters and
mse (Eq.~\ref{mseloss}) measures the accuracy of RNN.}
\label{tableY3}
{\small
\begin{tabular}{cccccccc}
\noalign{\smallskip} \hline \hline \noalign{\smallskip
}
{\tiny Model} & {\tiny $\text{RNN}_{n}$ } & {\tiny $\text{RNN}_{10n} $} & {\tiny $\text{LSTM}_{n} $} & {\tiny $\text{LSTM}_{10n} $} & {\tiny $\text{DR-RNN}_1 $} & {\tiny $\text{DR-RNN}_2$ } & {\tiny $\text{DR-RNN}_4$ } \\
\hline
$ d$ & 33 & 84 & 1093 & 4053 & 3 & 4 & 6\\
mse train & 33$ \cdot 10^{-2}$ & 17$ \cdot 10^{-2}$ & 33$ \cdot 10^{-2}$ &15 $\cdot 10^{-2}$ & 3$ \cdot 10^{-3}$ & 1$ \cdot 10^{-4}$ & 1$ \cdot 10^{-6}$ \\
mse test & 33$ \cdot 10^{-2}$ & 17$ \cdot 10^{-2}$ & 33$ \cdot 10^{-2}$ &15 $\cdot 10^{-2}$ & 4$ \cdot 10^{-2}$ & 5$ \cdot 10^{-4}$ & 1$ \cdot 10^{-6}$ \\
\noalign{\smallskip} \hline \noalign{\smallskip
}
\end{tabular} }
\end{table}
%%--------------------------------

%%--------------------------------
\begin{figure}[!h]
\centering
\includegraphics[width=\linewidth]{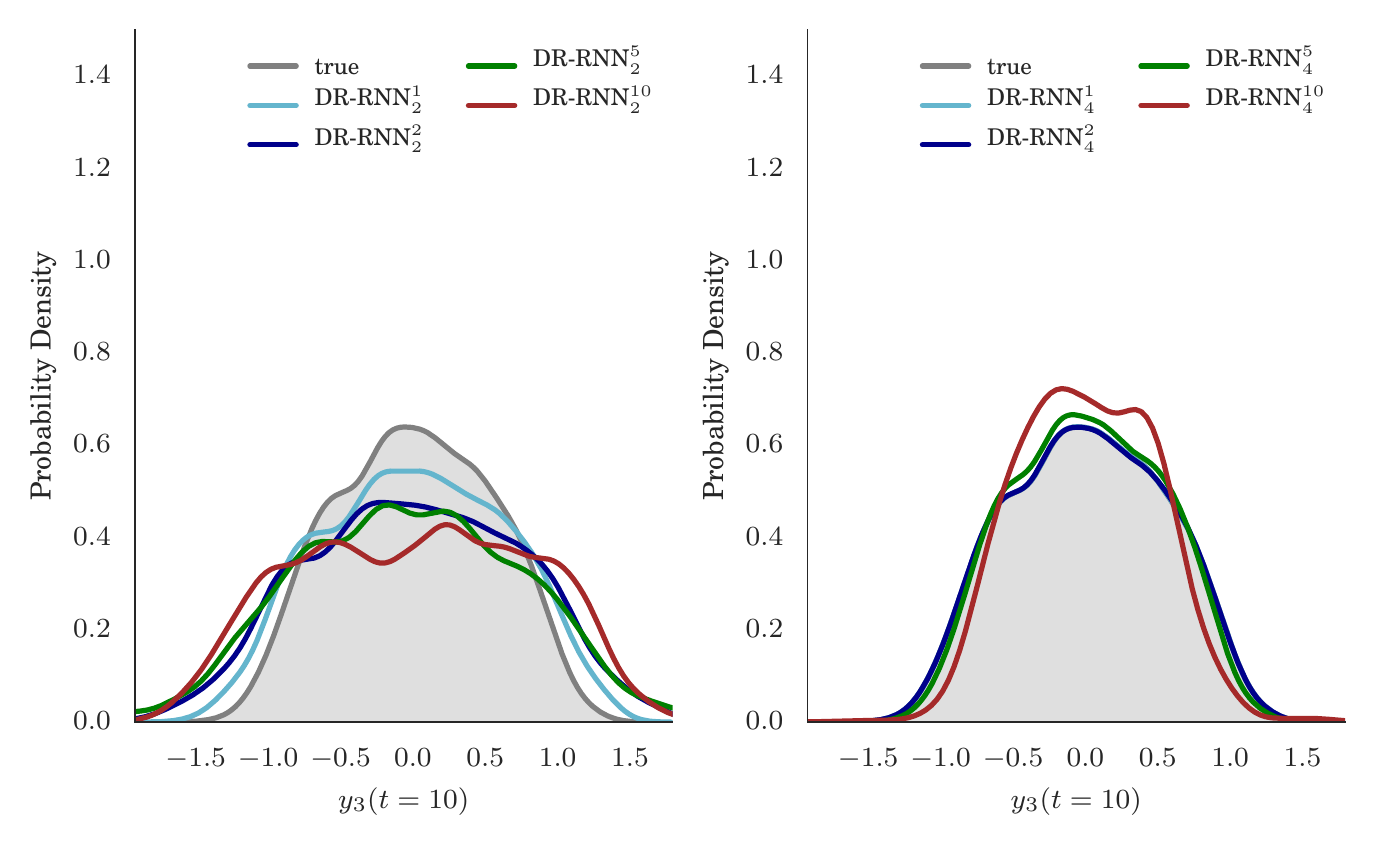}
\caption{Comparison of kernel density estimated probability density function (PDF) of $y_2(t=10)$ (left) and $y_3(t=10)$ (right) obtained from DR-RNN computed for different large time step size w.r.t. true PDF computed from fine step size in problem 3. Subscript in the label DR-RNN denotes the number of layers $K$ in DR-RNN. Superscript in the label DR-RNN denotes how many times the large time step bigger than the fine step size.}
\label{figY1Y2Y3largetimestep}
\end{figure}

%%--------------------------------
We carry out the same large time step performance analysis carried out in problem 1 and problem 2 for DR-RNN$ _2$ and DR-RNN$ _4$.
Figure~\ref{figY1Y2Y3largetimestep} compares the PDF of $y_3(t=10)$ using DR-RNN$_2$ and DR-RNN$_4$ for different large time step with the PDF computed from the reference solution using the fine time step size. One can notice the performance trend of DR-RNN$_2$ and DR-RNN$_4$ are nearly similar to the trend noticed in problem 1 and problem 2 (Figure~\ref{figY1largetimestep} and Figure~\ref{figY1Y2largetimestep}). From the results presented in Figure~\ref{figY1Y2Y3largetimestep}, we observe that DR-RNN$ _4$ performs well for large time steps of 2, 5 times $\Delta t = 0.1$, however, it results in small errors in the PDF plot for the case of 10 times $\Delta t = 0.1$ in this problem. % compared to the problem 1 and problem 2.
%%--------------------------------
%%--------------------------------
\subsection{Dimensionality reduction in space}
\label{secspatialmodelreduction}
In this section, we evaluate the performance of DR-RNN  in spatial dimensional reduction by using DR-RNN to approximate the ROM derived from POD-Galerkin strategy. We compare DR-RNN with POD based ROM and POD-DEIM ROM to conduct two parametric uncertainty quantification problems involving time dependent partial differential equations. 

In the following test cases, the weight matrix $\mU$ of DR-RNN in Eq.~\ref{DR-RNNequation} is initialized randomly from a uniform distribution function $\mathtt{U [0.1, 0.5]}$.  The vector training parameter $ \vw $ in Eq.~\ref{DR-RNNequation} is initialized randomly from the white Gaussian distribution with its standard deviation fixed to 0.1. The scalar training parameters $\eta_k$ in Eq.~\ref{DR-RNNequation}  are initialized randomly from the uniform distribution $\mathtt{U [0.1, 0.4]}$. We set the hyperparameters $\zeta$, $\gamma$ in Eq.~\ref{rmspropDR-RNN}  to $0.9, 0.1$ respectively. 

\subsubsection*{Problem 4}

\begin{figure}[h!]
\vspace{-8pt}
\centering
\includegraphics{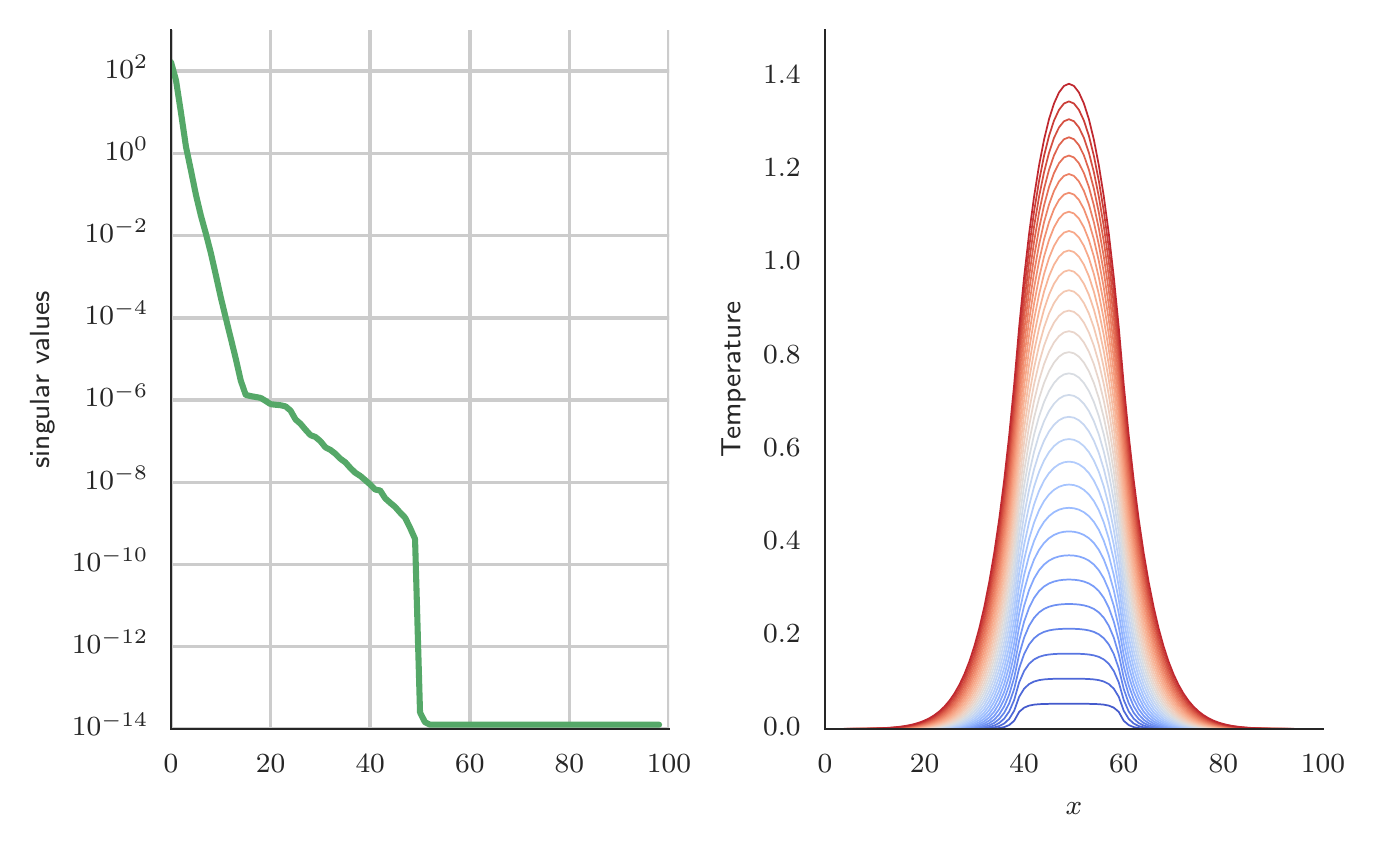}
\caption{Left: Singular values of the solution snapshot matrix $\mX$. Right: Numerical Solutions of the full-order system $n=99$ in problem 4.}
\label{temperatureprofilea}
\vspace{-8pt}
\end{figure}

In this problem we model unsteady heat diffusion over the spatial domain $x = [0, 1]$ using:
\begin{equation}
\dfrac{\partial \vy}{\partial t} = -\alpha~\dfrac{\partial^2 \vy }{\partial x^2} + \vg
\label{heatsource}
\end{equation}
where $\vy$ is the temperature field, $\alpha$ is the random heat diffusion coefficient with uniform probability distribution function  $\mathtt{U [0.01, 0.08]}$. The problem is imposed with homogeneous initial condition $\vy(x, 0) = 0$ and Dirchelet boundary conditions $y(0, t) = 0$ and $y(1, t) = 0$. The heat source $\vg$ takes the form:
\begin{equation}
\begin{aligned}
\vg&=\left\{\!\!\!
\begin{array}{ll}
1 & \text{if $x \in [0.4, 0.6]$ }\\
0 &
\text{else}
\end{array}\right.
%\label{heatsource}
\end{aligned}
\end{equation}
We use a finite difference discretization with a spatial step size $\Delta x = 0.01$. The discretized FOM is formulated as:
\begin{equation}
\dfrac{d \vy}{d t} = \mA~\vy + \vb
\label{FOMheatsource}
\end{equation}
with $\mA \in \mathcal{R}^{n \times n}$ obtained using second order central difference stencil. The dimension of the problem is $n=99$. The resulting system of ODEs (Eq.~\ref{FOMheatsource}) is then solved by using standard implicit Euler method with a time step size $\Delta t = 0.03$ for 40 time steps. We solve the problem for 500 random samples of $\alpha$.
Further, a set of solution snapshots is collected to construct the POD basis by computing the following singular value decomposition
\begin{equation}
\begin{aligned}
\mX &= \mU~\Sigma~\mW^* \qquad \mU \in \mathcal{R}^{n \times n} \qquad \Sigma \in \mathcal{R}^{n \times N_s} \qquad \mW \in \mathcal{R}^{N_s \times N_s}
\end{aligned}
\end{equation}
where $\mX$ is the snapshot matrix of the sample solutions of Eq.~\ref{FOMheatsource}, $N_s$ is the number of snapshots used in computing SVD. The space of $\vy$ is spanned by the orthonormal column vectors of matrix $\mU$. The left panel in the Figure~\ref{temperatureprofilea} shows the decay of singular values of the snapshot matrix $\mX$. The optimal basis for approximating $\vy(t)$ is given by the first $r$ columns of matrix $\mU$ denoted by $\mU_r$ and is used to reduce the FOM given by Eq.~\ref{FOMheatsource} to POD based ROM of the form:
\begin{equation}
\begin{aligned}
%\dfrac{d\vy^{t+1}}{dt} = \mA \cdot \vf(\vs^{t+1}) + \vb
%\dfrac{d \vy}{dt} = \mU_r^T~\mA~\vy + \mU_r^T~\vb \\
\dfrac{d \tilde{\vy}}{dt} = \tilde{\mA}~\tilde{\vy} + \tilde{\vb}
\end{aligned}
\label{heatROM}
\end{equation}
where $\tilde{\mA} = \mU_r^\top~\mA~\mU_r$ and $\tilde{\vb} = \mU_r^\top~\vb $.
Next, we solve Eq.~\ref{heatROM}  using standard implicit Euler method with a time step of size $\Delta t = 0.03$ for 40 time steps using the same 500 random samples of $\alpha$ used in FOM (Eq.~\ref{FOMheatsource}). We solve Eq.~\ref{heatROM} for a set of different number of POD basis functions ($r=2, 4, 5, 7, 15$). 
Finally, we built DR-RNN with four layers to approximate the ROM defined in Eq.~\ref{heatROM}. We train DR-RNN using time snapshot solutions of Eq.~\ref{heatROM} collected for some random samples of heat diffusion coefficient.
%%%%----------------------------
%%%%-----------------------------
%%%%%-----------------------

\begin{figure}[h!]
%\vspace{-8pt}
\centering
\includegraphics{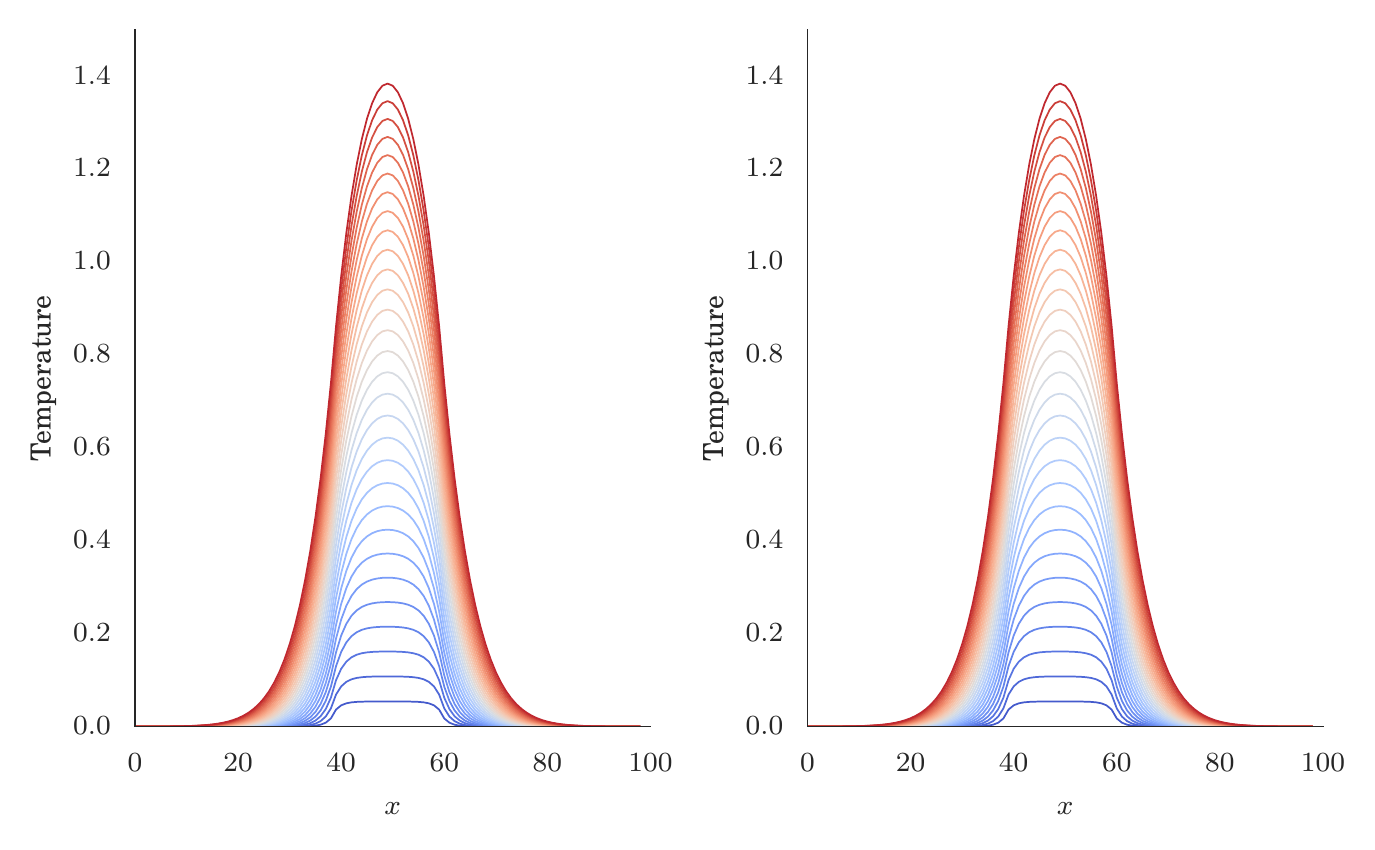}
\caption{Numerical solutions for problem 4 at different time steps. Left: POD-Galerkin reduced system with $15$ POD basis. Right: DR-RNN using $15$ POD basis. Dimension of the full-order model $n=99$.}
\label{temperatureprofileb}
\vspace{-8pt}
\end{figure}
%%%%-------------------------
Figures~\ref{temperatureprofilea} and~\ref{temperatureprofileb} show the numerical solutions obtained from the FOM, the linear POD of dimension 15, and the DR-RNN of dimension 15. The results plotted in the figures show that both the POD based ROM and the DR-RNN with dimension 15 produce good approximations to the original full-order system.

%%--------------------------------
\begin{figure}[!h]
\centering
\includegraphics{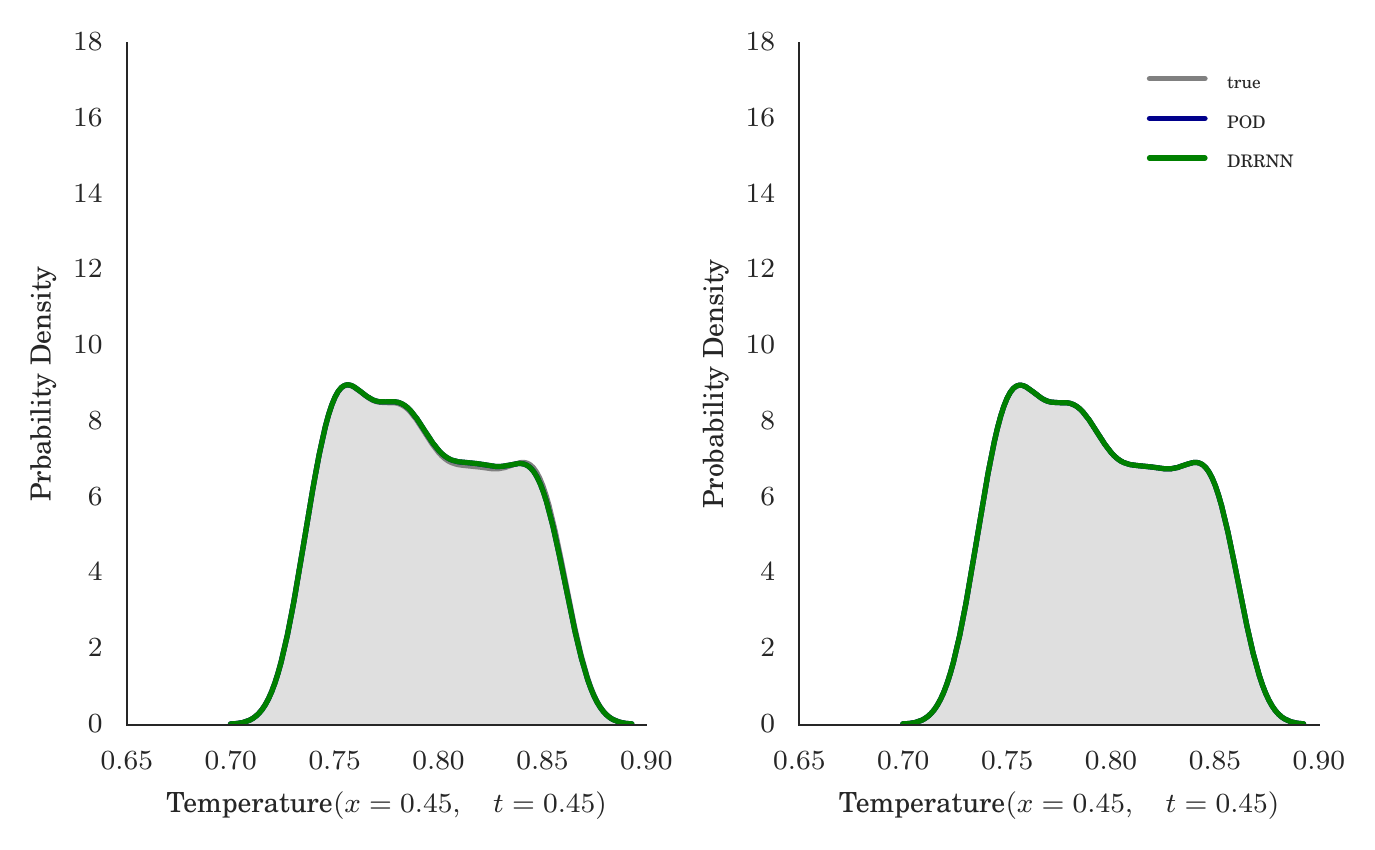}
\caption{Comparison of kernel density estimated probability density function (PDF) obtained from all ROM w.r.t. true PDF obtained from full-order system in problem 4. Left: number of POD basis used $=5$. Right: number of POD basis used $=15$. Dimension of the full-order model $n=99$.}
\label{figkdeplotheat}
\end{figure}
\begin{figure}[!h]
\centering
\includegraphics{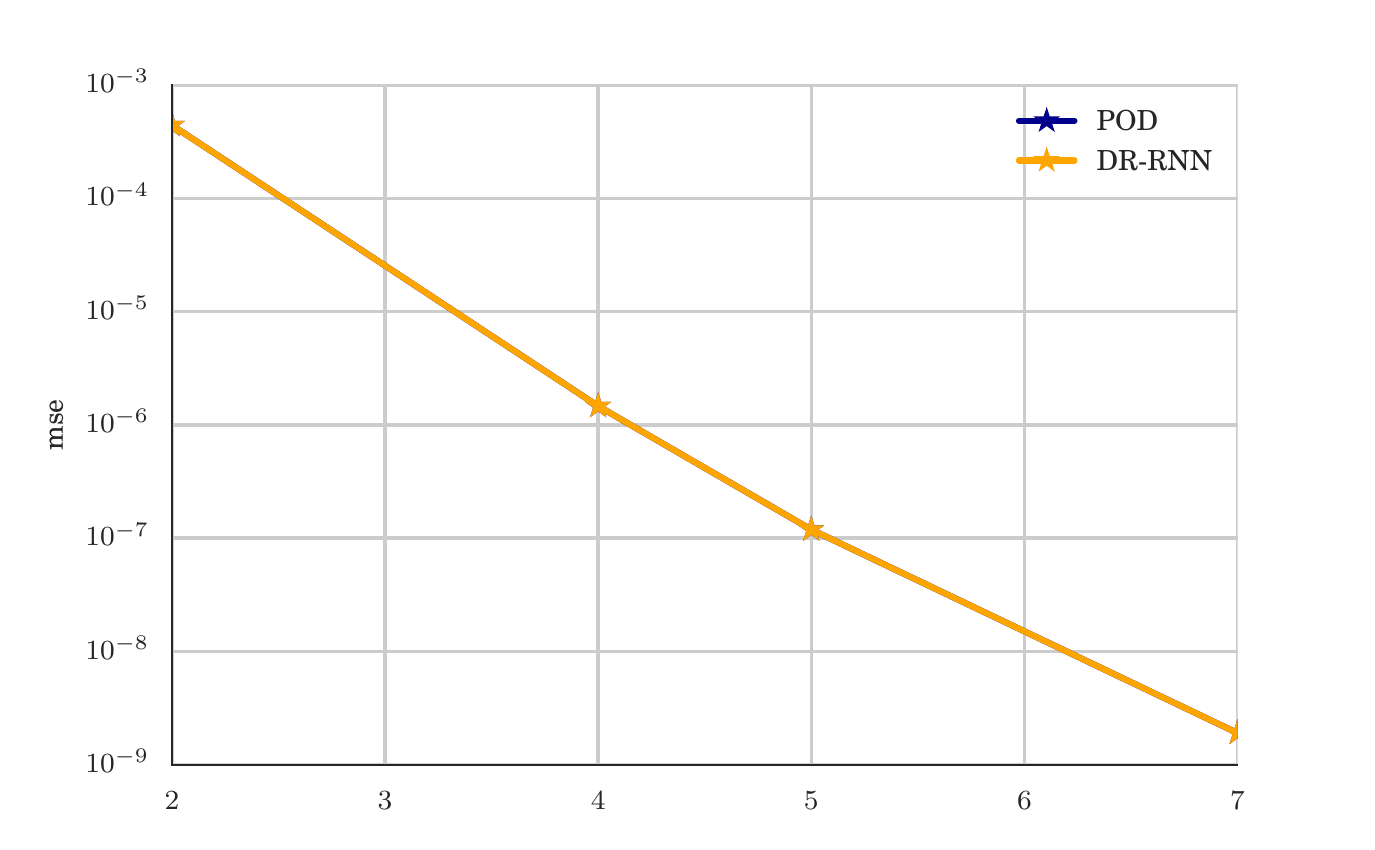}
\caption{Comparison of mse defined in Eq.~\ref{mseloss} obtained from POD and DR-RNN ROM in problem 4.}
\label{figtimeplotheatmse}
\end{figure}

%%--------------------------------
Figure~\ref{figkdeplotheat} compare PDF of $\vy(x=0.45, t=0.45)$ obtained from the reduced order models against the full order model PDF. The utilized ROMs use 5 POD basis functions in the left panel and 15 POD basis functions in the right panel. The results in Figure~\ref{figkdeplotheat} shows that the PDF obtained by the reduced systems are indistinguishable from the PDF of the FOM, while using $5$ or $15$ POD basis. Figure~\ref{figtimeplotheatmse} shows the mse defined in Eq.~\ref{mseloss} for different number of POD basis obtained from the POD based ROM and the DR-RNN. From the Figure~\ref{figtimeplotheatmse}, we can observe that the mse decreases with the increase in the number of POD basis due to the decay of singular values of the snapshot solution matrix $\mX_{\vs}$. Although the results of DR-RNN and POD based ROM are indistinguishable, we note that DR-RNN is an explicit method with a computational complexity of $\mathcal{O}(T \times L \times r^2)$ while POD method uses an implicit time discretization scheme with a complexity nearly to $\mathcal{O}(T \times L \times r^3)$, where $T$ is the number of time steps marched in the time domain and $L$ is the number of random samples.

%%--------------------------------
%%--------------------------------
\subsubsection*{Problem 5}
In this problem, we are interested in modeling the fluid displacement within a porous media, where water is pumped to displace oil. Although the displacing fluid is assumed to be immiscible with the displaced fluid (oil), the displacement front does not take place as a piston like flow process with a sharp interface between the two fluids. Rather, simultaneous flow of the immiscible fluids takes place within the porous media~\citep{chen2006computational}.
In this problem, we are mainly interested in the evolution of the saturation of the water phase. We solve a pair of partial differential equations namely the pressure and the saturation equations. A simplified one-dimensional pressure equation takes the form~\citep{chen2006computational}: 
\begin{equation}
\begin{aligned}
\dfrac{\partial}{\partial x} \left( \lambda \mK \cdot \dfrac {\partial \vp}{\partial x} \right)+ \vq = 0 \\
\end{aligned}
\label{pressure}
\end{equation}
where $\vp$ is the pressure, $\mK$ is the permeability, $\lambda$ is the total mobility, $\vq$ is the mass flow rate defined as $\vq = {\vq}_w / {\rho}_w + {\vq}_o / {\rho}_o$ and $\rho$ is the density. The subscript $w$ and $o$ denotes the water phase and the oil phase, respectively. The mobility term is defined as $\lambda = \lambda_w+\lambda_o$, where
$$\lambda_w = \frac{k_{rw}}{\mu_w} \qquad \lambda_o = \frac{k_{ro}}{\mu_o},$$
$\mu$ is the viscosity and $k_{rw}$, $k_{ro}$ are the relative permeability terms defined by the Brooks-Corey model~\citep{chen2006computational,aarnes2007introduction}. The second equation is the saturation equation defined as~\citep{chen2006computational}:
\begin{equation}
\begin{aligned}
\phi \dfrac{\partial \vs}{\partial t} + \dfrac{\partial~(\vv \cdot f_s)}{\partial x}
+ \frac{\vq_w}{\rho_w} = 0 
\end{aligned}
\label{saturation}
\end{equation}
where $\vv = -(\lambda_w+\lambda_o)~\mK~({\partial \vp}/{\partial x})$ is the Darcy velocity field, $\phi$ is the porosity and 
$f_s={\lambda_w} / (\lambda_w+\lambda_o)$ is the fractional flow function. We complete the model description by defining the phase relative permeabilities as a function of saturation using Brooks-Corey model~\citep{chen2006computational}:
\begin{equation}
\begin{aligned}
k_{rw} = {s^*}^{2} \qquad k_{ro} = {(1-{s}^*)}^2  \qquad  s^* = s - s_{or} - s_{ow}
\end{aligned}
%\label{saturation}
\end{equation}
where $s_{or}$ is the residual oil saturation, $s_{ow}$ is the residual water saturation and $s$ is the saturation value.  
In this simplified model, we assume a constant porosity throughout the media and we neglect the effects of compressibility, capillary, and gravity. We complete the description of the problem by the following input data:
\begin{equation}
\begin{aligned}
\vq(x=0) &= 0.1 \qquad \vq(x=1)=-0.1 \\
\mu_w&=0.1 \qquad \mu_o=1 \\
s_{or}&=0.2 \qquad s_{ow}=0.2
\end{aligned}
\end{equation}
The initial condition of $\vs$ is uniform and is equal to $s_{ow}$ and we use no flow boundary condition.
 We adopt a sequantial implicit solution strategy~\citep{chen2006computational,aarnes2007introduction} to compute the numerical solution of Eq.~\ref{pressure} and Eq.~\ref{saturation}.
In this method, a sequential updating of the
velocity field and saturation is performed where each equation is treated separately.
The first step is to solve for the pressure and the velocity field at an initial time. Then, with this velocity field and initial saturation, the saturation is evolved over a small number of time steps with the velocity field kept constant. The resulting saturation is then used to update the pressure and velocity. The process is repeated until the time of interest. We use a simple finite volume method (FVM) for spatial discretization with first order upwind scheme as it is a conservative method. The discretized form of the FOM of Eq.~\ref{saturation} is formulated as:
\begin{equation}
\begin{aligned}
%\dfrac{d\vs^{t+1}}{dt} = \mA \cdot \vf(\vs^{t+1}) + \vb
\dfrac{d \vs}{dt} = \mA~\vf(\vs) + \vb
\end{aligned}
\label{FOM}
\end{equation}

\begin{figure}[!h]
\centering %'firstkerasheatbigtime'+str(jumptime)+'deeplinearsourcesmalltruetimeconventional.pdf'
\includegraphics{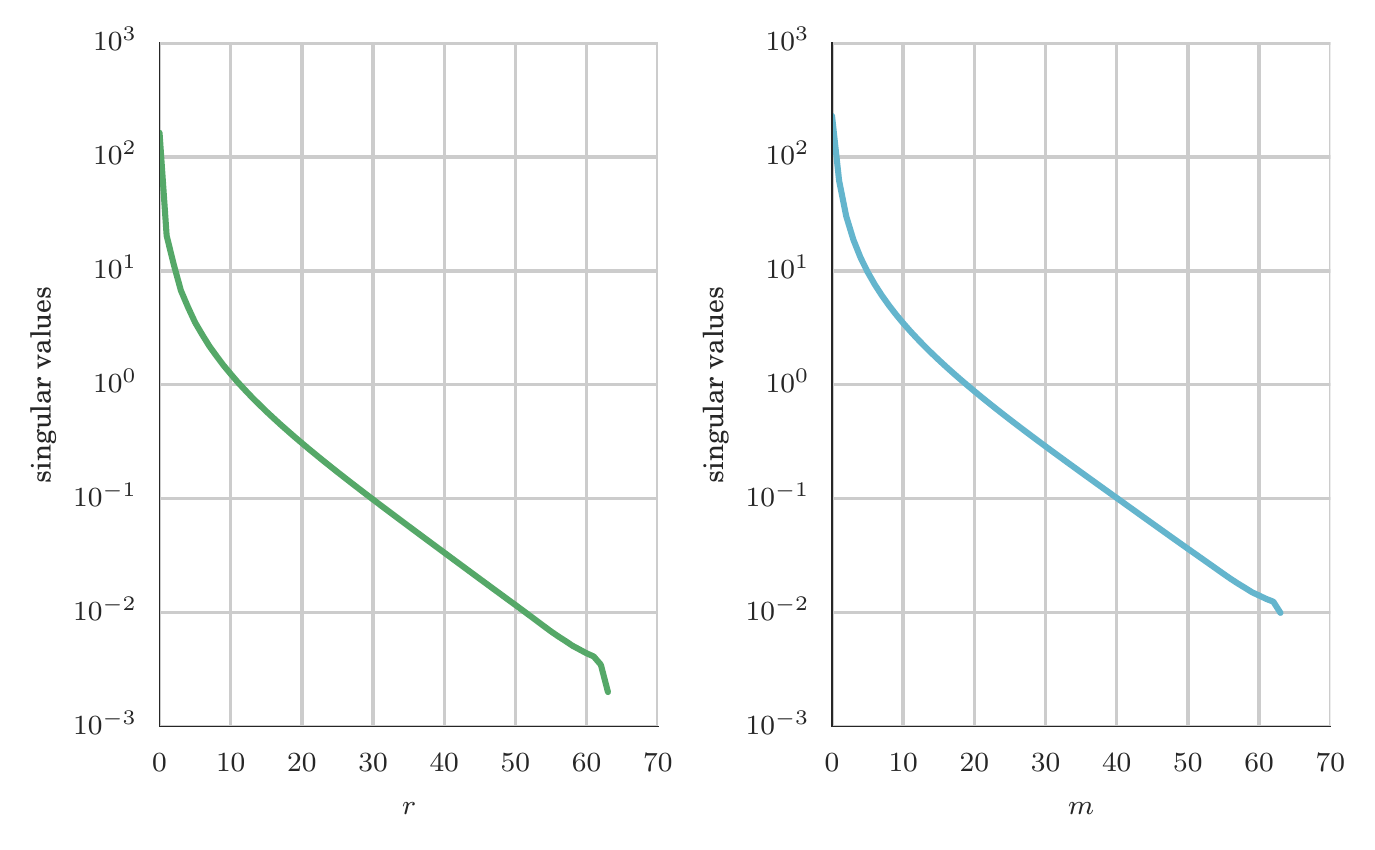}
\caption{Left: Singular values of the solution snapshot matrix $\mX_{\vs}$. Right: Singular values of the nonlinear function snapshot matrix $\mX_{\vf}$.}
\label{figsingularvalues}
\end{figure}

This equation is then discretized in time using backward Euler method. In space, we use 64 spatial grid points over the domain $x = [0, 1]$ and in time we use a time step of size  $\Delta t=0.015$ for 100  time steps. Newton Raphson iteration is used to solve the resulting system of nonlinear equations to evolve the saturation at each time step.  The uncertainty parameter in this test problem is the porosity value $\phi$ with a uniform probability distribution function 
$\mathtt{U [0.18, 0.38]}$.
We solve the FOM (Eq.~\ref{FOM}) by using standard implicit Euler method for 500 random samples of $\phi$. It is interesting fact to note that constant $\Delta t = 0.03$ violates Von Neumann stability condition given by 
\begin{equation}
\Delta t \le \frac{\phi \cdot {\Delta x}}{max \left( \vv \cdot \dfrac{df} {ds}\right)}
\label{VonNeumannstability}
\end{equation}
where $\Delta x$ is numerical grid size~\citep{pletcher2012computational, thomas2013numerical, aarnes2007introduction}. Next, the POD basis vectors are constructed from the solutions of the full-order system taken from the collected set of snapshot solutions. This is done by computing the following singular value decomposition
\begin{equation}
\begin{aligned}
\mX_{\vs} &= \mU~\Sigma_{\vs}~\mW_{\vs}^* \qquad \mU \in \mathcal{R}^{n \times n} \qquad \Sigma_{\vs} \in \mathcal{R}^{n \times N_s} \qquad \mW_{\vs} \in \mathcal{R}^{N_s \times N_s} \\
\mX_{\vf} &= \mV~\Sigma_{\vf}~\mW_{\vf}^* \qquad \mV \in \mathcal{R}^{n \times n} \qquad \Sigma_{\vf} \in \mathcal{R}^{n \times N_f} \qquad \mW_{\vf} \in \mathcal{R}^{N_f \times N_f}
\end{aligned}
\end{equation}
where $\mX_{\vs} $ is the snapshot matrix of saturation and $\mX_{\vf}$ is the snapshot matrix of nonlinear function $\vf(\vs)$, $n=64$ is the dimension of $\vs$ and $N_s$, $N_f$ are the number of snapshots used in computing SVD for the saturation and the nonlinear flow function respectively. The space of saturation is spanned by the orthonormal column vectors of matrix $\mU$ and the space of nonlinear function $\vf(\vs)$ is spanned by the orthonormal column vectors in the matrix $\mV$.  
The optimal basis for approximating $\vs(t)$ is given by the first $r$ columns of the matrix $\mU$ denoted by $\mU_r$ and is used to reduce FOM to POD based ROM of the form:
\begin{equation}
\begin{aligned}
%\dfrac{d\vs^{t+1}}{dt} = \mA \cdot \vf(\vs^{t+1}) + \vb
%\mU_r^\top~\dfrac{d \vs}{dt} = \mU_r^\top~\mA~\vf(\vs) + \mU_r^\top~\vb \\
\dfrac{d \tilde{\vs}}{dt} = \mU_r^\top~\mA~\vf(\mU_r~\tilde{\vs}) + \tilde{\vb}
\end{aligned}
\label{ROM}
\end{equation}
where $\vs \approx \mU_r~\tilde{\vs}$,  $\tilde{\vb}=\mU_r^\top~\vb $ and $\mU_r^\top~\mA~\vf( \mU_r~\tilde{\vs})$ forms the bottleneck that has to be reduced with DEIM as detailed in Section~\ref{secdeim}. Application of the DEIM algorithm (section~\ref{secdeim}) on nonlinear POD based ROM (Eq.~\ref{ROM}) results in POD-DEIM reduced order model of the form:
\begin{equation}
\begin{aligned}
%\dfrac{d\vs^{t+1}}{dt} = \mA \cdot \vf(\vs^{t+1}) + \vb
%\mU_r^\top~\dfrac{d \vs}{dt} = \mU_r^\top~\mA~\vf(\vs) + \mU_r^\top~\vb \\
\dfrac{d \tilde{\vs}}{dt} = \tilde{\mA}~\vf(\mP^\top \mU_r~\tilde{\vs}) + \tilde{\vb}
\end{aligned}
\label{ROMDEIM}
\end{equation}
where $\tilde{\mA} = \mU_r^\top~\mA~\mD$, $\mD= \mV_{m} (\mP^\top \mV_{m})^{-1}~$ referred as DEIM-matrix in Eq.~\ref{deimdetermined} (section~\ref{secdeim}), $\mV_{m}$ is the orthogonal matrix for optimally approximating $\vf(\vs)$  given by the first $m$ columns of the matrix $\mV$. Figure~\ref{figsingularvalues} shows the decay of singular values of the snapshot matrix $\mX_{\vs}$ and of the nonlinear function snapshot matrix $\mX_{\vf}$. Next, we solve Eq.~\ref{ROM} and Eq.~\ref{ROMDEIM}  by using standard implicit Euler method with a time step of $\Delta t = 0.03$ for 100 time steps using the same 500 random samples of $\phi$ used in FOM (Eq.~\ref{FOM}). We solve Eq.~\ref{ROM} for a set of POD basis functions ($r=15, 35, 55$) and similarly, we solve Eq.~\ref{ROMDEIM} for the same set of POD basis functions using a DEIM basis functions of fixed number ($m=35$). 
Further, we built DR-RNN to approximate the POD-DEIM ROM (Eq.~\ref{ROMDEIM}) where we apply DEIM in the DR-RNN to evaluate the nonlinearity, which gives an important speedup in the efficiency of the formulation.  We train DR-RNN using time snapshot solutions of Eq.~\ref{ROM} collected for some random samples of porosity values.
%
%%--------------------------------
%%--------------------------------

%%--------------------------------
\begin{figure}[!h]
\centering
\includegraphics{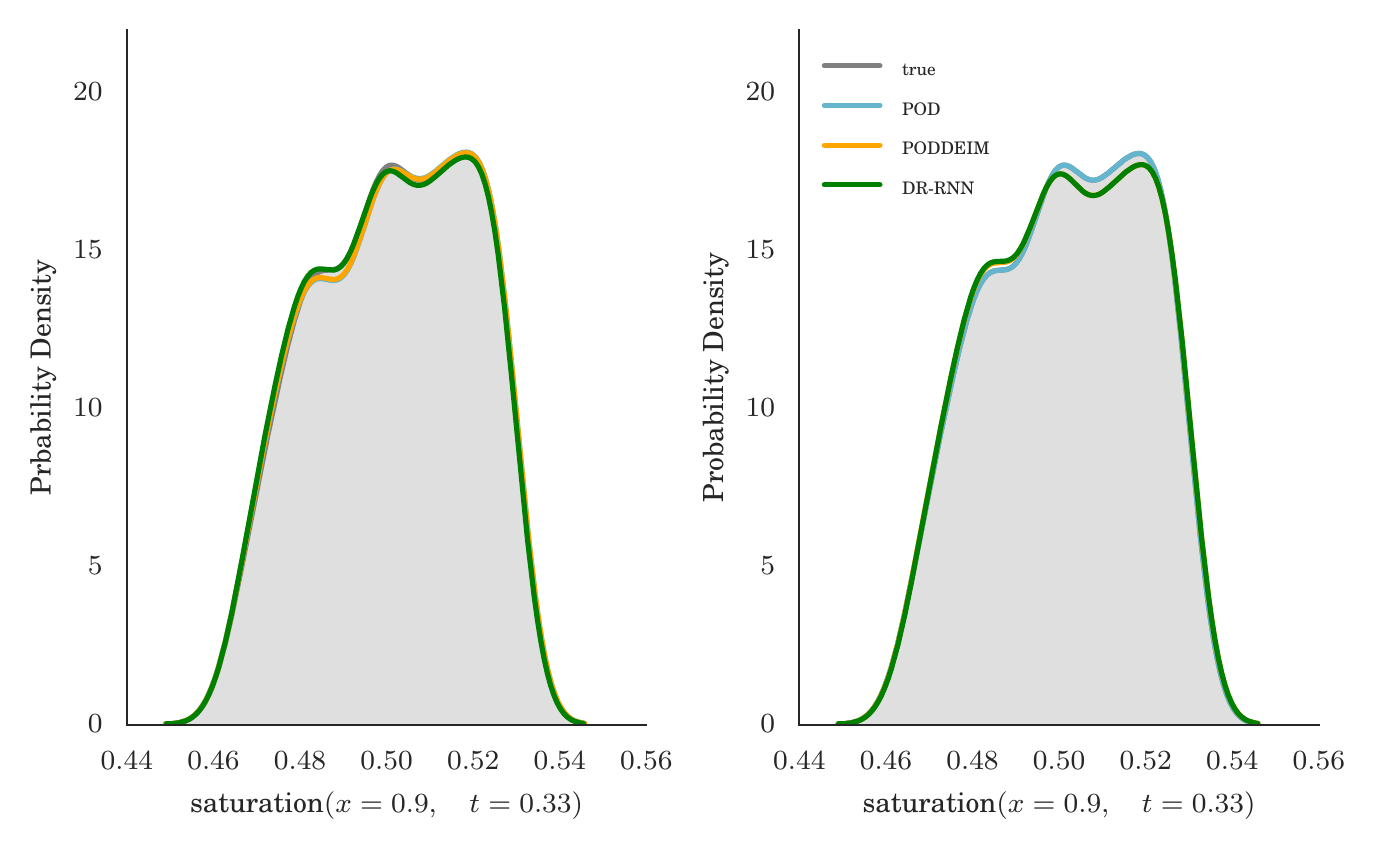}
\caption{Comparison of kernel density estimated probability density function (PDF) obtained from all ROMs w.r.t. true PDF obtained from full-order system in problem 5. Left: number of POD basis used $=15$. Right: number of POD basis used $=35$. Dimension of the full-order model $n=64$.}
\label{figkdeplotsaturation}
\end{figure}
%%--------------------------------

%%--------------------------------
\begin{figure}[!h]
\centering
\includegraphics{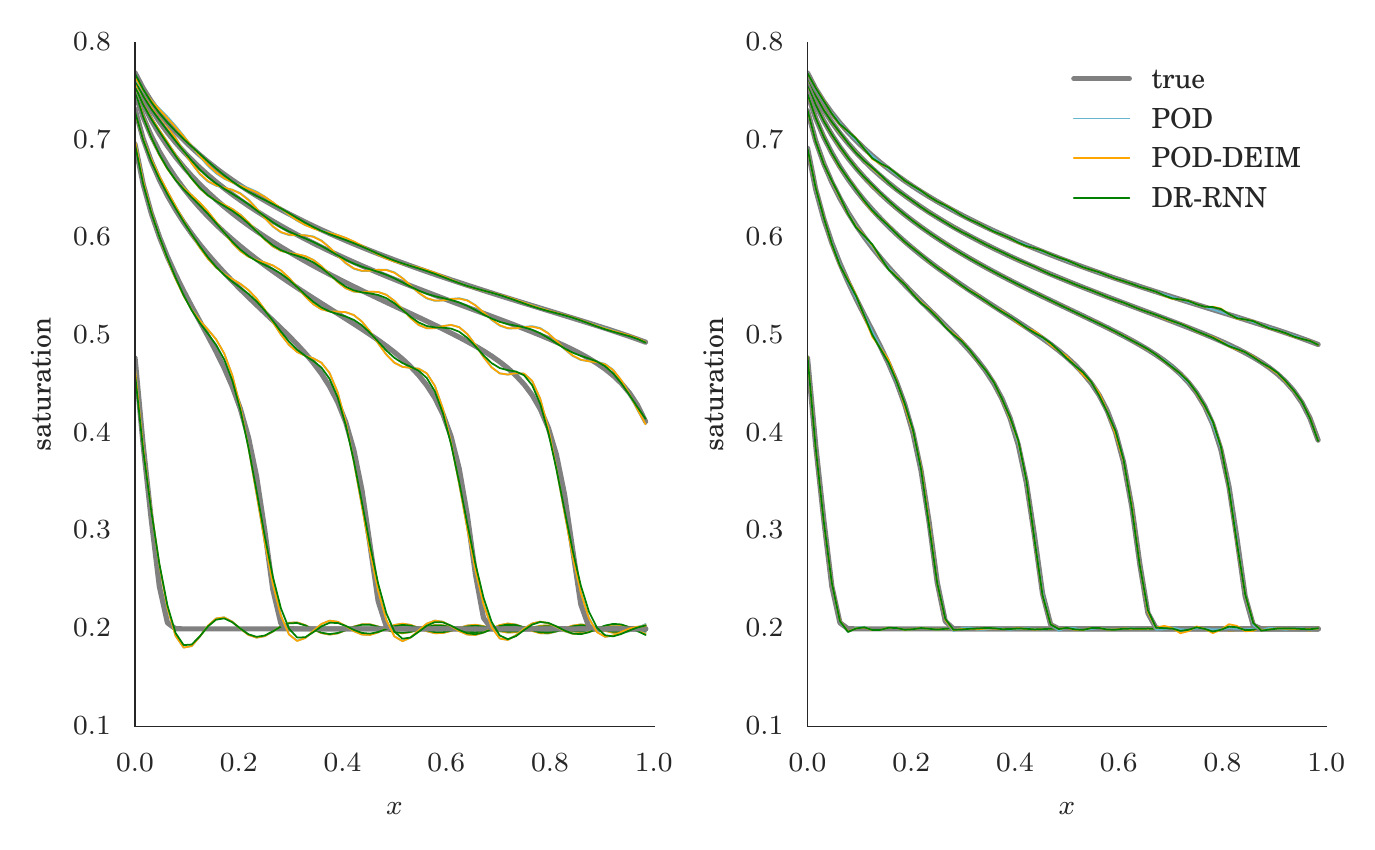}
\caption{Numerical solution of the saturation equation obtained from all ROMs w.r.t. full-order system in problem 5. Left: number of POD basis used $=15$. Right: number of POD basis used $=35$. Dimension of the full-order model $n=64$ and porosity value used $\phi=0.2$.}
\label{figtimeplotsaturation}
\end{figure}
%%--------------------------------

Figure~\ref{figkdeplotsaturation} compares the kernel density estimated probability density function (PDF) obtained from all ROMs to the PDF obtained from the FOM. Figure~\ref{figtimeplotsaturation} compares the numerical solutions obtained from all the reduced order models to the numerical solutions obtained from  the FOM. In these figures, ROM uses 15 POD basis functions in the left panel and 35 POD basis functions
in the right panel. From these figures, when the POD basis of dimension 35 is used, the numerical solutions of the reduced systems from all approaches appear to be indistinguishable from the numerical solution of the FOM. We note that the saturation equation has a hyperbolic structure which is more complicated to capture, especially in the nonlinear function.
%%--------------------------------
\begin{figure}[!h]
\centering
\includegraphics{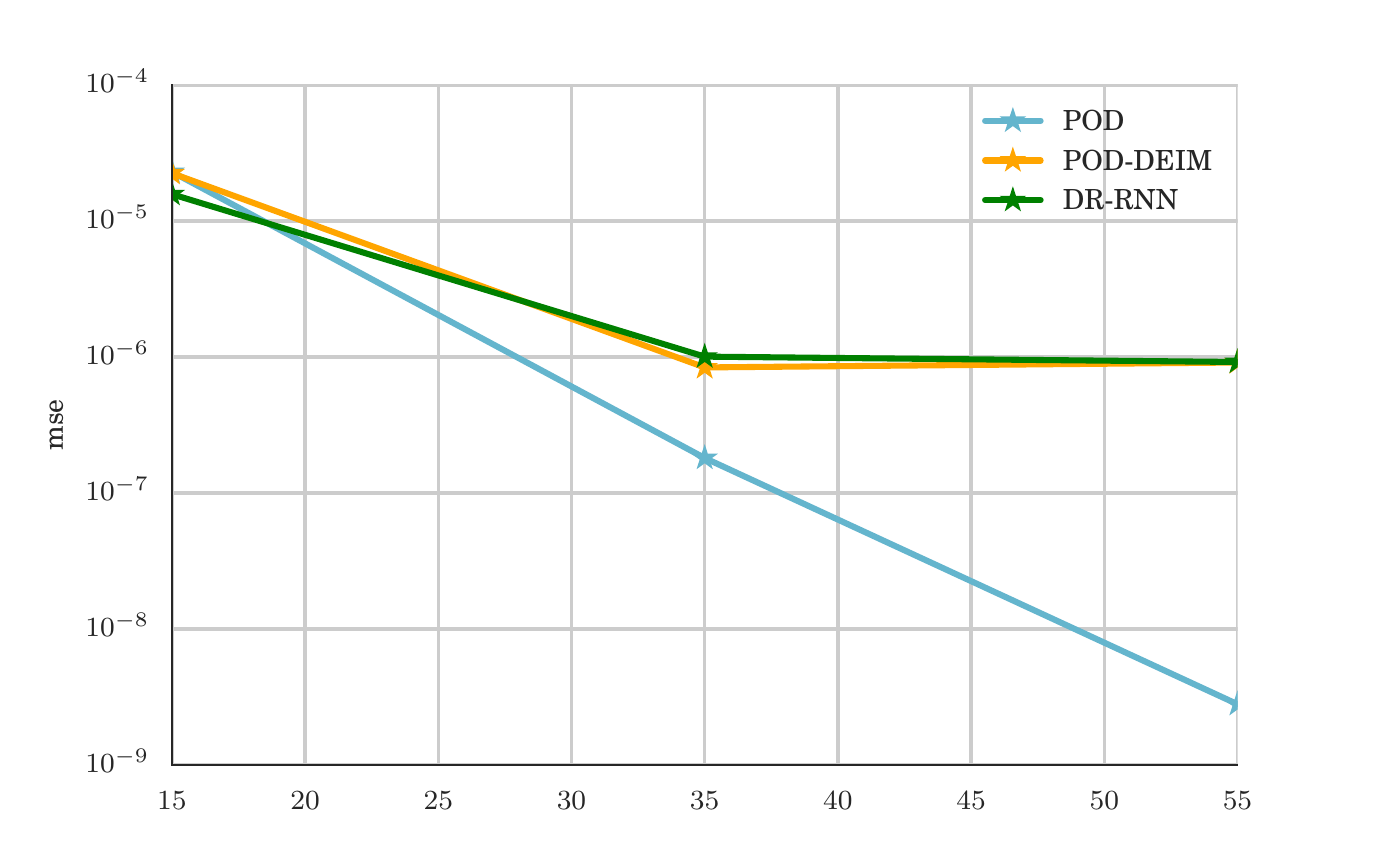}
\caption{Comparison of mse defined in Eq.~\ref{mseloss} obtained from all ROMs in problem 5.}
\label{figtimeplotsaturationmse}
\end{figure}
%%--------------------------------
Figure~\ref{figtimeplotsaturationmse} shows the mse defined in Eq.~\ref{mseloss} at different number of POD basis obtained from all ROMs.
From Figure~\ref{figtimeplotsaturationmse}, we can observe a decrease in mse as we increase the number of POD basis which is attributed to the decay of singular values of the snapshot solution matrix $\mX_{\vs}$. Although the errors from the POD reduced system is slightly lower than the errors arising from applying POD-DEIM and DR-RNN, the complexity in the on-line computation of the nonlinear function $\vf(\vs)$ still depends on the dimension of the original full-order system. Moreover, it is necessary to compute the Jacobian matrix of full dimension at every Newton iteration and at every time step in POD based ROM~\citep{deim}. Despite the fact that POD-DEIM approach not only gives an accurate reduced system with reduced computational complexity by removing the dependency on the dimension of the original full-order system with the general nonlinearities, POD-DEIM relies on evaluating the Jacobian matrix at every Newton iteration and results in a computational complexity in order $\mathcal{O}(T \times L \times p \times r^3)$, where $p$ is the number of Newton iterations. The presented numerical results showed that both POD-DEIM and DR-RNN approaches can be used to construct an accurate reduced system. 
However, DR-RNN constructs an accurate reduced system without evaluating the Jacobian matrix (as an explicit method) and thus limiting the computational complexity to $\mathcal{O}(T \times L \times K \times r^2)$ instead of $\mathcal{O}(T \times L \times p \times r^3)$, where $K \ll p$ is the number of stacked network layers and $p$ is the number of Newton iterations.
\section{Conclusions}
\label{secconclusion}
In this paper, we introduce a Deep Residual Recurrent Neural Network (DR-RNN) as an efficient model reduction technique that accounts for the dynamics of the full order physical system. We then present a new model order reduction for nonlinear dynamical systems using DR-RNN in combination with POD based MOR techniques. We demonstrate two different applications of the developed DR-RNN to reduce the computational complexity of the full-order system in different computational examples involving parametric uncertainty quantification. Our first application concerns the use of DR-RNN for reducing the computational complexity from $\mathcal{O}(n^3)$ to $\mathcal{O}(n^2)$ for nonlinear ODE systems. In this context, we evaluate DR-RNN in emulating nonlinear dynamical systems using a different number of large time step sizes violating the numerical stability condition for a small time discretization errors. 
The presented results show an increased accuracy of DR-RNN as the number of residual layer increases based on the hierarchical iterative update scheme.  

The second application of DR-RNN is related to spatial dimensionality reduction of dynamical systems governed by time dependent PDEs with parametric uncertainty. In this context, we use DR-RNN to approximate ROM derived from a POD-Galerkin strategy. For the nonlinear case, we combined POD with the DEIM algorithm for approximating the nonlinear function. The developed DR-RNN provides a significant reduction of the computational complexity of the extracted ROM limiting the computational complexity to $\mathcal{O}(K \times r^2)$ instead of $\mathcal{O}(p \times r^3)$ per time step for the nonlinear POD-DEIM method, where $K \ll p$ is the number of stacked network layers in DR-RNN and $p$ is the number of Newton iterations in POD-EIM.

%Possible future research directions are to use DR-RNN in realistic large scale problems, to train DR-RNN by minimising the norm of the nonlinear residual equations without training data. We also intent to use recent innovation in the machine learning namely deep reinforcement learning to more effectively construct approximations to the nonlinear dynamics using DR-RNN.

%\section*{References}
%\bibliographystyle{plainnat}
%\bibliography{myrefcorrected}

\end{document}